\documentclass[a4paper,fleqn,usenatbib]{mnras}

\usepackage[T1]{fontenc}
\usepackage{ae,aecompl}
\usepackage{color}

\usepackage{graphicx}	
\usepackage{amsmath}	
\usepackage{amssymb}	

\usepackage{times,txfonts}

\newcommand{\feh}{$\mathrm{[Fe/H]}$}
\newcommand{\afe}{$\mathrm{[\alpha/Fe]}$}
\newcommand{\alfe}{$\mathrm{[Al/Fe]}$}
\newcommand{\mgfe}{$\mathrm{[Mg/Fe]}$}
\newcommand{\nife}{$\mathrm{[Ni/Fe]}$}

\newbox\grsign \setbox\grsign=\hbox{$>$} \newdimen\grdimen \grdimen=\ht\grsign
\newbox\simlessbox \newbox\simgreatbox
\setbox\simgreatbox=\hbox{\raise.5ex\hbox{$>$}\llap
     {\lower.5ex\hbox{$\sim$}}}\ht1=\grdimen\dp1=0pt
\setbox\simlessbox=\hbox{\raise.5ex\hbox{$<$}\llap
     {\lower.5ex\hbox{$\sim$}}}\ht2=\grdimen\dp2=0pt
\def\simgreater{\mathrel{\copy\simgreatbox}}
\def\simless{\mathrel{\copy\simlessbox}}
\newbox\simppropto
\setbox\simppropto=\hbox{\raise.5ex\hbox{$\sim$}\llap
     {\lower.5ex\hbox{$\propto$}}}\ht2=\grdimen\dp2=0pt

\title[The origin of accreted stellar halo populations]{The origin of accreted stellar
halo populations in the Milky Way using APOGEE, \emph{Gaia}, and the
EAGLE simulations}

\author[J. T. Mackereth et al.]{
J. Ted Mackereth,$^{1}$\thanks{E-mail: J.E.Mackereth@2011.ljmu.ac.uk (LJMU)}
Ricardo P. Schiavon$^{1}$,
Joel Pfeffer$^{1}$,
Christian R. Hayes$^{2}$,
\newauthor
Jo Bovy$^{3}$,
Borja Anguiano$^{2}$,
Carlos Allende Prieto$^{4,5}$,
Sten Hasselquist$^{6}$,
\newauthor
Jon Holtzman$^{7}$,
Jennifer A. Johnson$^{8}$,
Steven R. Majewski$^{2}$,
Robert O'Connell$^{2}$,
\newauthor
Matthew Shetrone$^{9}$,
Patricia B. Tissera$^{10}$,
and J. G. Fern\'andez-Trincado$^{11,12,13}$
\\
\\
$^{1}$ Astrophysics Research Institute, Liverpool John Moores University, 146 Brownlow Hill, Liverpool, L3 5RF, UK \\
$^{2}$ Department of Astronomy, University of Virginia, Charlottesville, VA 22904-4325, USA \\
$^{3}$ Department of Astronomy and Astrophysics, University of Toronto, 50 St. George Street, Toronto, ON M5S 3H4, Canada \\
$^{4}$ Instituto de Astrofisica de Canarias, Via Lactea, 38205 La Laguna, Tenerife, Spain \\
$^{5}$Universidad de La Laguna, Departamento de Astrof\'isica, E-38206 La Laguna, Tenerife, Spain \\
$^{6}$ Department of Physics \& Astronomy, University of Utah, Salt Lake City, UT 84112, USA \\
$^{7}$ New Mexico State University, Las Cruces, NM 88003, USA \\
$^{8}$ Department of Astronomy and Center for Cosmology and AstroParticle Physics, Ohio State University, Columbus, OH, USA \\
$^{9}$ University of Texas at Austin, McDonald Observatory, Fort Davis, TX 79734, USA \\
$^{10}$ Department of Physics, Universidad Andres Bello, 700 Fernandez Concha, Chile \\
$^{11}$Departamento de Astronom\'\i{}a, Casilla 160-C, Universidad de Concepci\'on, Concepci\'on, Chile\\
$^{12}$ Instituto de Astronom\'ia y Ciencias Planetarias, Universidad de Atacama, Copayapu 485, Copiap\'o, Chile.\\
$^{13}$ Institut Utinam, CNRS UMR 6213, Universit\'e Bourgogne-Franche-Comt\'e, OSU THETA Franche-Comt\'e, Observatoire de Besan\c{c}on,\\ BP 1615, 25010 Besan\c{c}on Cedex, France
}
\date{Accepted XXX. Received YYY; in original form ZZZ}

\pubyear{2015}

\begin{document}
\label{firstpage}
\pagerange{\pageref{firstpage}--\pageref{lastpage}}
\maketitle

\begin{abstract}

Recent work indicates that the nearby Galactic halo is
dominated by the debris from a major accretion event.  We confirm
that result from an analysis of APOGEE-DR14 element abundances and
\textit{Gaia}-DR2 kinematics of halo stars.  We show that $\sim$~2/3
of nearby halo stars have high orbital eccentricities ($e \gtrsim
0.8$), and abundance patterns typical of massive Milky Way dwarf
galaxy satellites today, characterised by relatively low [Fe/H],
[Mg/Fe], [Al/Fe], and [Ni/Fe].  The trend followed by high $e$ stars
in the [Mg/Fe]-[Fe/H] plane shows a change of slope at [Fe/H]$\sim-1.3$,
which is also typical of stellar populations from relatively massive
dwarf galaxies. Low~$e$ stars exhibit no such change of slope within
the observed [Fe/H] range and show slightly higher abundances of
Mg, Al and Ni.  Unlike their low~$e$ counterparts, high~$e$ stars
show slightly retrograde motion, make higher vertical excursions
and reach larger apocentre radii.  By comparing the position in \mgfe{}-\feh{} space of
high~$e$ stars with those of accreted galaxies from the EAGLE suite
of cosmological simulations we constrain the mass of the accreted
satellite to be in the range $10^{8.5}\lesssim M_*\lesssim
10^{9}\mathrm{M_\odot}$. We show that the median orbital eccentricities
of debris are largely unchanged since merger time, implying that
this accretion event likely happened at $z\lesssim1.5$.  The exact
nature of the low $e$ population is unclear, but we hypothesise
that it is a combination of {\it in situ} star formation, high $|z|$
disc stars, lower mass accretion events, and contamination by the
low $e$ tail of the high $e$ population.  Finally, our results imply
that the accretion history of the Milky Way was quite unusual.

\end{abstract}
\begin{keywords}
Galaxy: formation -- Galaxy: halo -- Galaxy: kinematics and dynamics
-- Galaxy: stellar content -- Galaxy: abundances
\end{keywords}



\section{Introduction} \label{intro}

It is now well established that accretion of lower mass systems is
a fundamental component of the evolution and mass build up of
galaxies \citep{1991ApJ...379...52W}.  Due to its very long dynamical
timescale, the stellar halo of the Milky Way keeps a record of the
Galaxy's past accretion activity.  That record can be accessed
through the collection of precision 6D phase space and multi-element
abundance information for very large samples of halo stars, which
together enable fundamental tests of galaxy formation models.  While
this field has a long history
\citep[e.g.,][]{1962ApJ...136..748E,1978ApJ...225..357S}, we highlight
only a few of the main contributions from the past decade, for
brevity.  \citet{2010A&A...511L..10N} and \citet{2012A&A...538A..21S}
were the first to identify the presence of an older, high~$\alpha$
and a younger, low~$\alpha$ halo population at metallicity lower
than that of the Galactic disc in the solar neighbourhood.
The kinematics of those stellar populations suggested an {\it in
situ} \footnote{ By stars formed {\it in situ} we mean those
that were formed within the Galaxy, either from gas originally
associated with its dark matter halo or that which was accreted
onto it.} or accreted origin, respectively. More
recently, \citet{2015MNRAS.453..758H} proposed abundance ratio
diagnostics to distinguish accreted from {\it in situ} halo stars,
arguing that the accreted population dominates the nearby halo.
Using APOGEE data, \citet{2018ApJ...852...50F} and
\citet{2018ApJ...852...49H} studied the chemical compositions and
kinematics of the metal-rich nearby halo, suggesting that much of
the low \mgfe{} halo population is associated with the debris of
accreted satellites, likely with a similar star formation history
to the Large Magellanic Cloud (LMC).

Studies of the Galactic halo are being revolutionised by the advent
of large astrometric, photometric, and spectroscopic surveys of the
stellar populations of the Galaxy.  The \emph{Gaia} astrometric
satellite has opened new avenues for exploration of substructure
in phase space, with the potential for new discoveries further
amplified by the addition of chemical information from spectroscopic
surveys.  Indeed, combining \emph{Gaia} parallaxes and proper motions
\citep{2016arXiv160904303L,2018arXiv180409365G} with spectroscopic
data from SDSS \citep{2000AJ....120.1579Y} and APOGEE
\citep{2015arXiv150905420M}, two groups have identified what seems
to be the accretion of a relatively massive stellar system that
dominates the stellar populations of the nearby halo.  Analysing a
sample of SDSS-\emph{Gaia} DR1 main sequence stars,
\citet{2018MNRAS.478..611B} showed that the velocity ellipsoid of
halo stars becomes strongly anisotropic for stars with [Fe/H]$>$--1.7.
Comparing their data to a suite of N-body only cosmological
numerical simulations, they concluded that such orbital configurations
are likely to result from the accretion of a massive satellite at
about the time of the formation of the Galactic disc, roughly between
$z=1$ and 3.

Based on \emph{Gaia} DR2 data \citep{2018arXiv180409365G},
\citet{2018arXiv180500453M} determined the configuration of MW
globular clusters (GCs) in action space.  They find that 12 GCs in
the halo are consistent with an origin in a single massive accretion
event, consistent with the conclusions reached by
\citet{2018MNRAS.478..611B}.  \citet{2018arXiv180500453M} find that
these clusters have highly eccentric orbits, at $e \gtrsim 0.85$,
and suggest that the fact that all the clusters occupy a similar
region in action space supports the idea that this highly anisotropic
stellar population in the halo is mainly formed from the debris of
a single accretion event.

In a follow up study, \citet{2018arXiv180510288D} estimated the
orbital parameters for a sample of nearby main-sequence and distant
horizontal-branch stars by combining \emph{Gaia} DR2 data
with spectroscopic outputs from SDSS-DR9 \citep{2012ApJS..203...21A}.
They found that the apocentre radii of a significant population of
stars in the halo appear to ``pile up'' at an $r_\mathrm{ap} \sim 20$
kpc. The authors link this population with that found by
\citet{2018MNRAS.478..611B}.   
This result has special significance in light of the analysis of
numerical simulations by \citet{2013ApJ...763..113D}, who proposed
that the existence of a ``break radius'' in the Milky Way halo,
beyond which the stellar density drops precipitously, is associated
with the ``pile up'' of stellar apocenters at a comparable
Galactocentric distance.  \citet{2013ApJ...763..113D} argue that
the observed existence of a break radius in the Milky Way halo and
the absence of such a break in the Andromeda galaxy (M31) suggests
that the latter had a much more prolonged accretion history than
the former.  Follow up work using the same sample suggests
that, inside this break radius, roughly 50\% of the halo is made
up of stars from this accretion event \citep{2018arXiv180704290L}.

An independent analysis of \emph{Gaia} DR2 data conducted by
\citet{2018ApJ...860L..11K} identified the presence of a large,
old, and metal-poor slightly counter-rotating structure in phase space.
They concluded that this population is associated with a relatively
massive object which, they hypothesised, may have been responsible
for the heating of the thick disc.  Following up on that result,
\citet{2018arXiv180606038H} used kinematic, chemical, and age
information for a large sample of stars in \emph{Gaia} and APOGEE
to identify a population of metal-poor stars with the same phase
space characteristics as those reported by
\citet{2018ApJ...860L..11K}.  The distribution of this stellar
population in the \afe{}-\feh{} plane, with relatively low \afe{}
and a large spread in \feh{}, suggests the chemical evolution trend
of a relatively massive system\footnote{The recent paper by
\citet{2018arXiv180707269F} hints at a similar conclusion, also on
the basis of APOGEE data.}.  Moreover, the positions of the stars
in the HR diagram are consistent with old ages (10-13 Gyr).  According
to  \citet{2018arXiv180606038H}, the accretion of a dwarf galaxy
with a mass similar to that of the Small Magellanic Cloud
\citep[see also][]{2018ApJ...852...49H} $\sim 10$ Gyr ago may have
been responsible for the heating of the thick disc. The notion that
the thick disc was formed from the vertical heating of a thinner
progenitor disc competes
with the so-called ``upside-down'' formation scenario \citep[see,
e.g,][]{2013ApJ...773...43B,2017arXiv170901040N} according to which
the early gaseous disc was thick as a result of strong stellar
feedback \citep[and/or clumpy gas accretion, e.g.,][]{2004ApJ...612..894B},
and slowly settled as the star formation waned to form the thinner
components of the disc.  Despite their differences, both scenarios
are consistent with at least some heating of the stellar disc by
satellite mergers, which in turn are also likely necessary to explain
the flaring of high \afe{} mono-age disc populations
\citep[e.g.][]{2015ApJ...804L...9M,2017arXiv170600018M}.  In general,
recent observational results do not seem to point towards a scenario
where the thick disc formed thin and was heated entirely by mergers,
which would produce a plateau in the age or $\alpha$~abundance
against scale height relationship that is not currently borne out
by the data
\citep[e.g.][]{2012ApJ...751..131B,2012ApJ...753..148B,2012ApJ...755..115B,2016ApJ...823...30B,2016MNRAS.455..987C,2017arXiv170600018M}

It is also important to note that the same population was identified
by \citet{2018ApJ...863..113H}, who studied Gaia DR2 and APOGEE-based
colour-magnitude diagrams, kinematics, and chemistry to identify a
population of metal-poor stars with high transversal velocities and
typically low or retrogade rotation, which they associate with the
last significant merger undergone by the Milky Way.

In addition to, and in support of these findings,
\citet{2018MNRAS.tmp.1537K} recently inferred that the Milky Way
has had a rather atypical assembly history given its mass, based
on analysis of the age-metallicity relation of Galactic GCs. They
found that the assembly rate of the Milky Way was among the uppermost
quartile of galaxies in their simulation, and identified three
recent massive accretion events. Those authors proposed that two of these accretion events correspond to the Sagittarius
dwarf and Canis Major\footnote{It was acknowledged by \citet{2018MNRAS.tmp.1537K} that `Canis Major' as it is known to the community is no longer considered as a genuine accreted stellar population. The term is used there historically, as many of the GCs were those originally associated with `Canis Major' and so, for clarity, we adopt it here when discussing those results.}. The most massive of those
accretion events is suggested to have no known debris, and have a
stellar mass $> 10^9 \mathrm{M_\odot}$, and correspond to GCs that
reside close to the Galactic center.  It indeed may be
possible, on the basis of the analysis of the Galactic GC population
by \citet{2018MNRAS.tmp.1537K}, that the accreted satellite
identified by \citet{2018MNRAS.478..611B} and \citet{2018ApJ...860L..11K} is associated with Canis Major, given that the GCs
identified by \citet{2018arXiv180500453M} are further out in the
halo, and those potentially associated with Canis Major
are located at Galactocentric distances $> 10$ kpc.  The finding
that the assembly history of the Milky Way is atypical is also
consistent with the work of \citet{2018MNRAS.477.5072M}, who found
that Milky Way stellar mass galaxies in the EAGLE simulation with
\afe{} abundance patterns similar to the Milky Way had
atypical accretion histories, characterised by early, rapid
accretion, which slowed at late times.

In summary, the local stellar halo has been shown to be dominated
by a population of moderately metal-poor, low~\afe{}, old stars on
highly eccentric orbits.  This population is the likely remnant of
a major accretion event that took place at about the same time that
the Galactic disc was itself forming.  These results have important
implications, which prompted us to examine the chemical and kinematic
properties of the newly discovered stellar population in detail.
In this paper we present an analysis of the
 abundance pattern of stars in common between the \emph{Gaia}-DR2
and APOGEE-DR14 catalogs, and discuss the implications of their
kinematic properties in light of the  EAGLE suite of numerical
cosmological simulations. We extend the studies of the element
abundances in these populations to include odd-$Z$ and Iron peak
elements, and examine the detailed kinematics of stars in sub-populations
defined by abundances and orbit eccentricity. We also extend previous
theoretical work on this population by examining the kinematics of
accreted debris from a fully self-consistent cosmological simulation
that provides a cosmologicallly motivated sample of accreted satellite
debris onto Milky Way mass haloes.  In Section \ref{data} we describe
our sample selection and orbital parameter determination, as well
as the details of the EAGLE simulations.  In Section \ref{highe}
we discuss the chemical and kinematic properties of this population.
In Section \ref{eagle} we contrast the kinematic properties of the
newly discovered stellar population with the expectations from
cosmological numerical simulations.  Our conclusions are summarised
in Section \ref{finale}.

\section{Sample and Data} \label{data}
\subsection{APOGEE DR14}

Our study is based on a cross-match between the SDSS-APOGEE DR14
and \emph{Gaia} DR2 catalogues. APOGEE \citep{2015arXiv150905420M}
is a near infrared spectroscopic survey of the stellar populations
of the Galaxy and its Local group neighbours. The data employed in
this paper come from the DR14 catalogue \citep{2018ApJS..235...42A},
which comprises a re-reduction and analysis of APOGEE-1 data
\citep[from SDSS-III,][]{2011AJ....142...72E}, alongside a set of
newly reduced and analysed observations from APOGEE-2 \citep[taken
as part of SDSS-IV,][]{2017AJ....154...28B}.  APOGEE DR14 contains
high S/N, R$\sim$22,500 spectra, radial velocities, stellar
photospheric parameters, and element abundances for over 270,000
stars in the $H$-band (1.5-1.7$\mathrm{\mu m}$).   Observations are
carried out using the 2.5m SDSS Telescope at Apache Point Observatory
(APO) \citep{2006AJ....131.2332G}, and fibre-fed to the APOGEE
spectrograph \citep{2010SPIE.7735E..1CW}. Targeting is performed
so as to simplify as much as possible the survey selection function
whilst preferentially selecting red giant stars, by employing
selection bins in the apparent $H$-band magnitude, and a simple
colour selection in dereddened $(J-K)_0$
\citep{2013AJ....146...81Z,2017AJ....154..198Z}. Spectra are reduced,
combined and then analysed through the APOGEE data reduction pipeline
\citep{2015AJ....150..173N}, and the APOGEE Stellar Parameters and
Chemical Abundances Pipeline \citep[ASPCAP,][]{2016AJ....151..144G}.
ASPCAP relies on a pre-computed library of synthetic stellar spectra
\citep{2015AJ....149..181Z} computed using a customised linelist
\citep{2015ApJS..221...24S} to measure stellar parameters, 19 element
abundances and heliocentric radial velocities of MW stars
\citep{2015AJ....150..148H}. Abundances are well tested against
samples from the literature \citep[][in press.]{jonssondr14}.  We
use distances for stars in APOGEE DR14 measured by the Brazilian
Participation Group \citep[BPG,][]{2016A&A...585A..42S}, included
in a publicly available Value-Added Catalogue (VAC). These distances
are measured using an early version of the StarHorse code
\citep{2018MNRAS.476.2556Q}, and combine spectroscopic and photometric
information to make a Bayesian distance estimation. The precision
of these measurements is expected to be $\sim 15 \%$, which we
determine to be similar to distance estimates derived from the
current \emph{Gaia} parallaxes at the range of distances spanned
by our sample of interest.  Given that the parallax measurements
can be uncertain in some cases, and especially so for distant stars,
the use of spectro-photometric distances is well motivated.

We examined the distribution of our sample stars in chemical
composition space considering all elemental abundances available
in the APOGEE DR14 catalog.  However, in this paper we choose to
focus on the abundances of Fe, Mg, Al, and Ni, which are the ones
providing interesting insights into the nature of the accreted halo
stellar population. The elemental abundances are measured as part
of the ASPCAP pipeline, which uses a two-step process. First, the
stellar parameters $T_{\mathrm{eff}}, \log(g), v_\mu, \mathrm{[M/H]},
\mathrm{[\alpha/M]}, \mathrm{[C/M]},$ and $\mathrm{[N/M]}$ (where
$v_\mu$ is the micro-turbulent velocity) are determined via a global
fit to the aforementioned spectral library \citep{2015AJ....149..181Z}.
The individual element abundances are then calculated by adjusting
the $\mathrm{[M/H]}$ ($\mathrm{[C/M]}$ and $\mathrm{[N/M]}$ for
Carbon and Nitrogen, and $\mathrm{[\alpha/M]}$ for $\alpha$ elements)
of the best-fit spectrum, and finding the best match to the observed
spectrum in windows around features in the spectrum which are
dominated by each element. The abundances are then all estimated
consistently, and can then be calibrated internally relative to
open cluster observations. The internal calibrations are performed
to account for systematic abundance variations with $T_\mathrm{eff}$.
In DR14, an external calibration is applied that forces the abundance
ratios of solar metallicity stars located near the solar circle to
be equal to solar \citep[][in press]{holtzdr14}.  This small
zero-point correction should be taken into consideration when making
comparisons between our results and other data.

\subsection{\emph{Gaia} DR2 and cross matching}

The ESA-\emph{Gaia} mission is a space-based astrometric survey
which is providing an unprecendented mapping of MW stars in phase
space. The second data-release, \emph{Gaia} DR2
\citep{2018arXiv180409365G}, provides 5-parameter astrometry (proper
motions, positions and parallaxes) for over 1.3 million objects in
the Galaxy.  Combined with accurate radial velocities and
spectro-photometric distance estimates from APOGEE DR14, these data
make possible the calculation of 6D phase space coordinates for
objects in common between the surveys.  Many improvements were made to the
data-processing between \emph{Gaia} DR1 \citep{2016arXiv160904303L}
and DR2, examples of which include: improvements to the source
detection algorithm, better modelling of the spacecraft attitude,
and the fact that DR2 uses its own reference frame based on quasars
(whereas DR1 was tied to the \emph{Tycho-2} and \textsc{Hipparcos}
catalogues for proper motion measurements). As a result of these
improvements, the typical uncertainty on astrometric parameters
is expected to be $\sim 0.2$ to $0.3$~mas in the middle of the
magnitude range (going up to $\sim 2$~mas for the faintest sources).
While the exact selection function of \emph{Gaia} is as yet not
well known, DR2 has improved completeness in bright stars, and the
survey is expected to be complete between $G=12$ and 17.

We perform a cross-match between APOGEE DR14 and \emph{Gaia} DR2
using the CDS X-match
service\footnote{\url{http://cdsxmatch.u-strasbg.fr/xmatch}} and
adopting a conservative position mismatch tolerance of 0.5$"$.  We
find that the full, uncut APOGEE
DR14 catalogue has 254,789 matched objects in \emph{Gaia} DR2 ($\sim
99\%$), 83,189 of which have full 6D phase-space coordinates (using
APOGEE radial velocities), have no warning or bad flags from the
APOGEE reduction and ASPCAP analysis, and were not observed during
commissioning of the APOGEE instrument (the main factor
that reduces the sample size are the APOGEE data quality flag cuts).
Of these objects, 81,491 have reliable distance measurements in the
APOGEE DR14 distance VAC.  We transform the observed data into the
Galactocentric coordinate frame, assuming the solar motion of
\citet{2010MNRAS.403.1829S}, propagating the observational uncertainties
while accounting for the correlation between errors in the \emph{Gaia}
data. The sample extends from Galactocentric cylindrical radii $R
\sim 3$~kpc out to $ R > 15$~kpc, reaching up to a maximum of
$10$~kpc away from the midplane. Throughout the paper, we assume
the solar radius $R_0 = 8$~kpc, and its distance from the midplane
$z_0 = 0.025$~kpc. The combined spectra of our final sample
have a minimum SNR $\sim 40$, and a median SNR $\sim 150$, corresponding
to median uncertainties on \feh{} $\sim 0.01$ dex, \mgfe{} $\sim
0.02$ dex, \alfe{} $\sim 0.05$ dex, and \nife{} $\sim 0.02$ dex.

Stars located within $r = 3\ r_{\mathrm{tidal}}$ of the centres of
known globular clusters were excluded from the sample.  Tidal radii,
$r_\mathrm{tidal}$, and cluster centres were adopted from the 2010
edition of the \citet{1996AJ....112.1487H}
catalog\footnote{\url{http://physwww.mcmaster.ca/~harris/mwgc.dat}}.  This
conservative cut removes 19 stars, after removal of stars not
belonging to the main APOGEE sample (e.g., stars from APOGEE ancillary
science programs).

\subsection{Orbital parameters}

To study the kinematical structure of the halo population in the
APOGEE-\emph{Gaia} catalogue, it is necessary to estimate the orbital
parameters of the stars and their associated uncertainties robustly.
To make these estimates, we use the fast orbit parameter estimation
method of \citet{2018arXiv180202592M}, which adapts the St\"ackel
fudge method for estimating action-angle coordinates in axisymmetric
potentials \citep[presented in][]{2012MNRAS.426.1324B} to directly
estimate the orbital eccentricity, $e$, apo- and pericentre radii,
$r_\mathrm{ap}$ and $r_\mathrm{peri}$, and the maximum vertical
excursion, $Z_\mathrm{max}$, to high precision and without recourse
to orbit integration (which can make the proper propagation of
uncertainties computationally costly at this scale).  We also
estimate the orbital actions, $J_{R}, L_{Z},$ and $J_{Z}$ for each
star via the same method. All estimates are performed using the
implementation of the St\"ackel approximation in the python package
\texttt{galpy} \citep{2015ApJS..216...29B}, and assuming the
\texttt{MWPotential2014} Milky-Way mass model included in \texttt{galpy}.
For each star, we Monte-Carlo (MC) sample the errors by constructing
the covariance matrix of the observed data.  We sample 100 realisations
of the observed coordinates of each star, and compute the orbit
parameters and actions for each sampled point, cataloguing the
median value of the samples for each parameter, their standard
deviation, and the correlation between parameters.  Performing the
orbital parameter estimation with many more than 100 samples makes
little difference to the median and standard deviation obtained,
so this number was selected for computational efficiency.

\subsection{The EAGLE simulations} 
\label{eagledata} 

In Section \ref{eagle} we undertake a simple analysis of simulated
galaxies from the EAGLE suite
\citep{2015MNRAS.446..521S,2015MNRAS.450.1937C}.  EAGLE models the
formation and evolution of galaxies in the context of $\mathrm{\Lambda
CDM}$ cosmology. The simulations are run using a version of the
smoothed particle hydrodynamics (SPH) and TreePM gravity solver
\textsc{Gadget 3} \citep[described most recently by
][]{2005MNRAS.364.1105S}, modified to include the
\citet{2013MNRAS.428.2840H} pressure-entropy formulation of SPH, a
time-step limiter \citep{2012MNRAS.419..465D}, and switches for
artificial viscosity and conduction \citep[as proposed
by][]{2010MNRAS.408..669C,2010MNRAS.401.1475P}. EAGLE produces a
realistic population of galaxies by virtue of the calibration of
its feedback efficiency to the observed galaxy stellar mass function
(GSMF), the stellar vs. black hole mass relation and galaxy disc
sizes, but has been shown to reproduce a broad range of observed
galaxy properties and scaling relations, such as the Tully-Fisher
relation \citep{2017MNRAS.464.4736F}, colour-magnitude relationships
\citep{2015MNRAS.452.2879T,2016MNRAS.460.3925T,2017MNRAS.470..771T},
and galaxy size evolution \citep{2017MNRAS.465..722F}. All particle
data and galaxy catalogues are now available publicly
\citep[see][]{2016A&C....15...72M} and are accessible
online\footnote{\url{http://galaxy-catalogue.dur.ac.uk}}.

The EAGLE suite offers various box sizes ranging from 12.5 to 100
cMpc on a side. Here, we examine the higher resolution simulation
adopting the `Recalibrated' model parameters
\citep[see][]{2015MNRAS.446..521S}, referred to here as L025N752-Recal
(we use the lower resolution, `Reference' model simulations to test
the numerical convergence of our results in Appendix \ref{sec:appB}).
The motivation behind the use of the higher resolution simulation
is the better resolution of small dwarf galaxies, and therefore
better sampling of the accreted galaxy population within the EAGLE
haloes. L025N752-Recal has dark matter particles of mass $1.21\times
10^{6}\ \mathrm{M_{\odot}}$ and an initially equal number of SPH
particles with mass $2.26\times 10^{5}\ \mathrm{M_{\odot}}$, adopting
a Plummer-equivalent gravitational softening length
$\epsilon_\mathrm{com}=1.33$ ckpc, limited to a maximum proper
length of $\epsilon_\mathrm{prop}=0.35$pkpc. We select a sample of
galaxies which have virial masses at $z=0$ roughly equal
to that proposed for the Milky Way, between $M_{200} = $ 0.8 and
2.0 $ \times 10^{12}\mathrm{M_{\odot}}$.
In the L025N752-Recal simulation, this corresponds to $N=22$ galaxies
with a wide range of stellar masses, and a range of assembly and
accretion histories.  The mean halo mass of the galaxies is
$0.9\times10^{12}\ \mathrm{M_\odot}$.  The majority (81\%) of the
galaxies are disk dominated systems, as defined using the methodology
of \citet{2017arXiv170406283C}, by calculating the fraction of
kinetic energy of a galaxies stars invested in ordered co-rotation
in the plane perpendicular to the vertical component of
the angular momentum, $\kappa_{\mathrm{co}}$.  Disk dominated
galaxies are defined as those with $\kappa_{\mathrm{co}} > 0.4$.
The remaining 19\% are a combination of galaxies undergoing mergers
and those without significant disk components. We do not cut the
sample based on this information, as we are interested in the debris
accreted in the lifetime of the galaxies, and not their eventual
morphology. Moreover, we show in Section \ref{eagle} that
our general conclusions are upheld regardless of the $z=0$ galaxy
morphology.  This sample makes a good testbed for understanding the
origin of satellite debris in the Milky Way, allowing us to study
the $z=0$ characteristics of a diverse sample of satellite galaxies
accreted onto Milky-Way-like haloes.

To calculate the eccentricities for the simulated star particles,
we solve (using the bisection method) the equation
\begin{equation}
L^2 + 2 r^2 [\Phi(r) - E] = 0 ,
\end{equation}
where L is the angular momentum, $\Phi(r)$ the gravitational potential
and E the total energy of the particle. For a bound orbit the
equation has two solutions (the peri- and apocentre distances, $r_\mathrm{peri}$
and $r_\mathrm{ap}$), which are equal for a perfectly circular orbit \citep[][eq.
3.14]{2008gady.book.....B}. Unbound particles are disregarded.  The
eccentricity is then calculated as
\begin{equation}
e = \frac{r_\mathrm{ap}-r_\mathrm{peri} }{r_\mathrm{ap}+r_\mathrm{peri} }.
\end{equation}

In practice, performing this estimation requires the assumption
that the potential is spherically symmetric, calculated using the
mass enclosed in the sphere with radius, $r$, which is equal
to the distance of the particle to the galaxy center. As most of
the accreted particles reside in the haloes of the central galaxies,
this provides a good approximation of the potential. Performing an
equivalent analysis to that performed on the data would require a
procedure for fitting an axisymmetric potential to EAGLE galaxies,
which is beyond the scope of this study. As the assumed potential
approaches spherical symmetry these two approximations become
equivalent, and so we are confident that the comparison is sound,
given that most of the APOGEE-\emph{Gaia} stars being considered
here reside in the Galactic halo (where the potential can be assumed
to be near-spherical).

\section{The high eccentricity halo population} \label{highe}

\begin{figure*}
\includegraphics[width=1.0\textwidth]{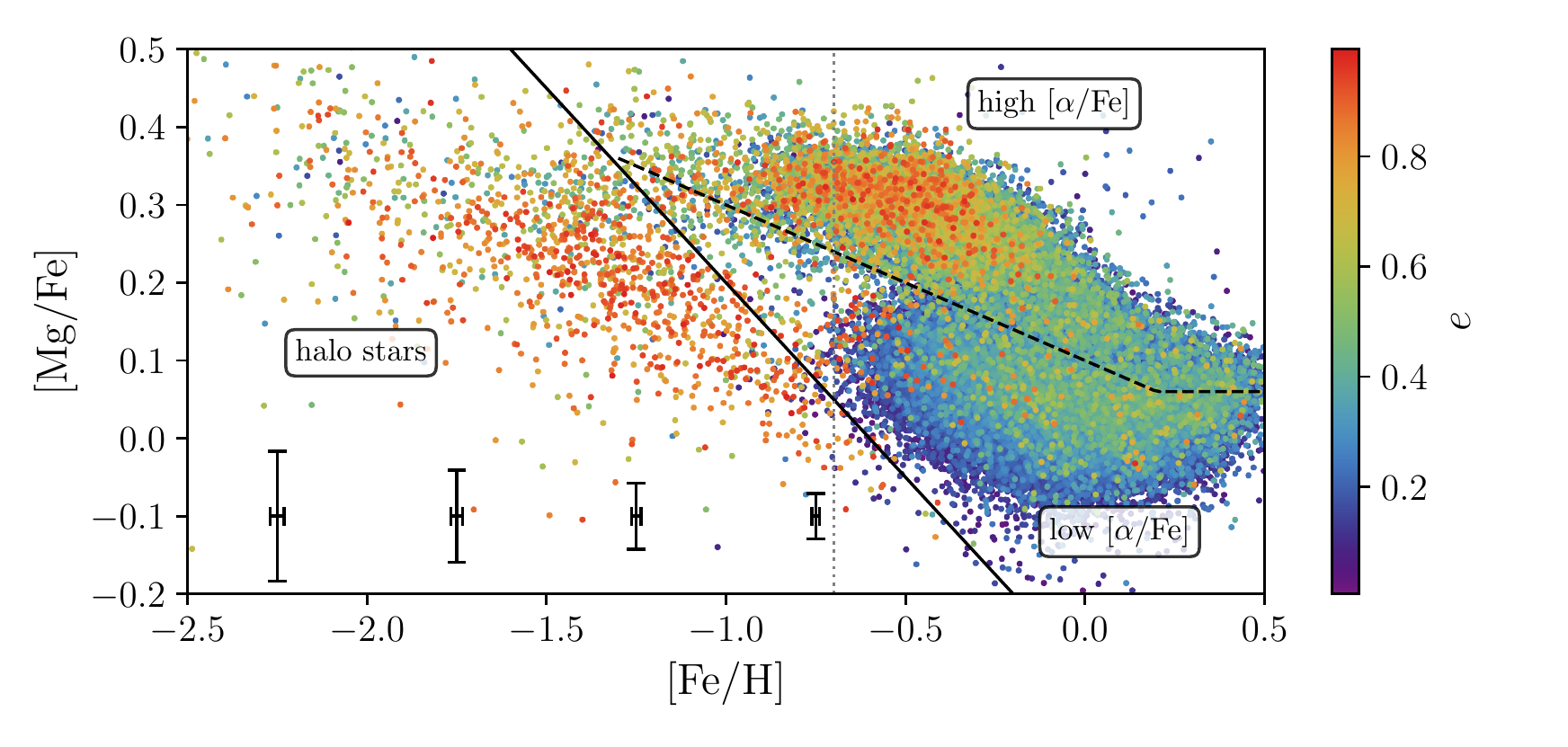}
\caption{\label{fig:characterisation} The \mgfe{}-\feh{} plane in
APOGEE DR14, coloured by orbital eccentricity $e$, as estimated
using the method of \citet{2018arXiv180202592M}. The points are
plotted with the highest $e$ stars overlaying the points at lower
$e$, such that the highest $e$ populations stand out.  Plotted
also are the mean error bars within \feh{} bins of width 0.5 dex
between -2.5 and -0.5 dex. It is clear that a population extends from
(\feh{},\mgfe{}) $\sim$  (-2.0, 0.3) to $\sim$ (-1.0, 0.1) that
appears to consist mainly of stars on highly eccentric orbits, with
a distinct element abundance pattern to that of the Galactic disc
(at \feh{}$>-0.7$). The dotted gray line reflects the cut in \feh{}
which is imposed to perform the $k$-means analysis.}
\end{figure*}

As mentioned above, \citet{2018arXiv180606038H} used APOGEE-DR14
abundances to show that a stellar population they initially identified
in phase space occupies a distinct locus in the \afe{}-\feh{} plane.
Among all $\alpha$ elements available from APOGEE spectra, 
Mg is the one for which abundances are the most reliable in the
metal-poor regime, due to the number of available lines and their
strength at low metallicity.  Therefore we decide to first look at
how stellar populations are distributed in the Mg-Fe plane as a
function of orbital eccentricity.  We show the \mgfe{}-\feh{}
distribution of APOGEE DR14 in Figure \ref{fig:characterisation},
where symbols are colour-coded according to each particle's orbital
eccentricity.

Three main groups of stars are apparent in \mgfe{}-\feh{} space,
which we demarcate using solid and dashed black lines (the
dotted gray line indicates the upper limit imposed on \feh{} for
the analysis in Section \ref{sec:abundances}).  The focus of this
paper is on the group we call \textit{halo stars} -- a term
we adopt merely as a label, which does not necessarily imply a
definitive assignation of every member star to the Galactic halo.
These stars occupy the same locus as the stellar population
identified by other groups as associated to a major accretion event
(see Section \ref{intro}), as well the ``LMg'' stars classed
as accreted halo by \citet{2018ApJ...852...49H} and
\citet{2018ApJ...852...50F} on the basis of APOGEE DR13 chemistry
and kinematics.  This group extends between (\feh{},
\mgfe{}) $\sim(-2.0, 0.3)$ to $\sim (-0.5, -0.1)$ and is dominated
by high eccentricity stars.  The other groups are the well known
high-$\alpha$ and low-$\alpha$ {\it disc} populations, characterised
in detail in several previous studies
\citep[e.g.][]{2014ApJ...796...38N,2015ApJ...808..132H,2015MNRAS.453.1855M,2017arXiv170600018M},
and commonly conflated with the thick and thin discs, respectively.
As expected, the latter populations are dominated by stars in very
circular orbits, although the high-$\alpha$ group contains a
non-negligible population of stars in fairly eccentric orbits 
($e > 0.7$) -- a topic that will be explored in a future study.

Careful inspection of the ``halo stars'' in
Figure~\ref{fig:characterisation} suggests that there is some
dependence of \mgfe{} on eccentricity, in that stars with higher
$e$ have slightly lower \mgfe{}.  Further examination of the data
suggests that the same is the case for \alfe{} and possibly \nife{}
and to a lesser extent [(C+N)/Fe].
 This finding motivates us to attempt an identification
of sub-structure in chemo-kinematic space in an objective and
data-driven fashion, so as to avoid ``cherry-picking'' arbitrary
selection limits in various parameters.  In this way we hope to
characterise the populations within the halo star locus by sub-dividing
its members into meaningful groups.  We use the \texttt{scikit-learn}
$k$-means clustering algorithm \citep{scikit-learn} in the space
of \feh{}, \mgfe{}, \alfe{}, \nife{} and eccentricity $e$.  The
choice of these parameters is guided by the fact that, on one hand,
these elements fall into the main groups: $\alpha$, odd-z and Fe
peak elements, respectively, and on the other hand it appears, at
least from Figure \ref{fig:characterisation}, that eccentricity is
a useful discriminator between disc and halo populations.  We limit
the maximum \feh{} to --0.7, to minimise contamination by disc
stars, finding that high $e$ stars in the low \afe{} disc locus at
higher \feh{} are not chemically similar to those in the ``halo
stars'' population and can be confidently disregarded. We set the
assumed number of clusters at $k= 4$, to anticipate the expected
separation of the high and low \afe{} disc, and then to allow for
subdivision of the halo population into any potentially meaningful
groups.  We find that the algorithm groups the high and low \afe{}
disc stars separately, and finds two clear groups in the lower
\feh{} space.  Setting $k>4$ subdivides the lower \feh{} groups in
an unstable manner, while $k=4$ provides very good stability over
many iterations of the algorithm. We also test the algorithm at $k
< 4$, finding that the stability is decreased also at lower $k$.
If the ``halo stars'' group from Figure \ref{fig:characterisation}
is isolated, using $k=2$ clustering can still recover the two groups
found when clustering the whole data set. We retain the full data-set
and use the $k=4$ clustering to avoid resorting to an arbitrary
selection of the division between halo and disc stars. Re-scaling
the data and re-running the clustering algorithm leads to
negligible differences in the results. It is worth noticing that
both halo populations have lower abundance ratios than thick disc
stars (in the region where they overlap in [Fe/H]) for C+N, Si, K,
Ca, and possibly also Mn.

\begin{figure}
\includegraphics[width=1.0\columnwidth]{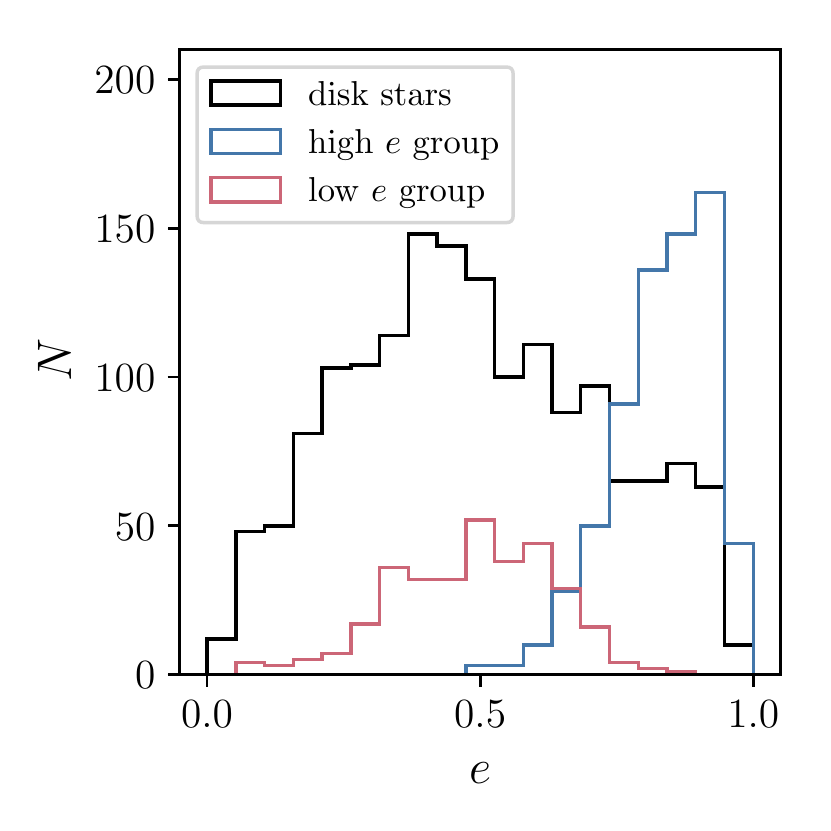}
\caption{\label{fig:eccdist} Eccentricity distributions of stars
in three groups identified by performing a $k$-means detection of
structure in the eccentricity-\feh{}-\mgfe{}-\alfe{}-\nife{} space.
The $k$-means algorithm cleanly separates the accreted halo component
into two groups, one characterised by low eccentricities (in red),
and the other with a peak at very high $e$ (in blue). We show the
stars assigned to the disc (with \feh{} $< -0.7$) in black as a
comparison sample. }
\end{figure}

\begin{figure}
\includegraphics[width=1.0\columnwidth]{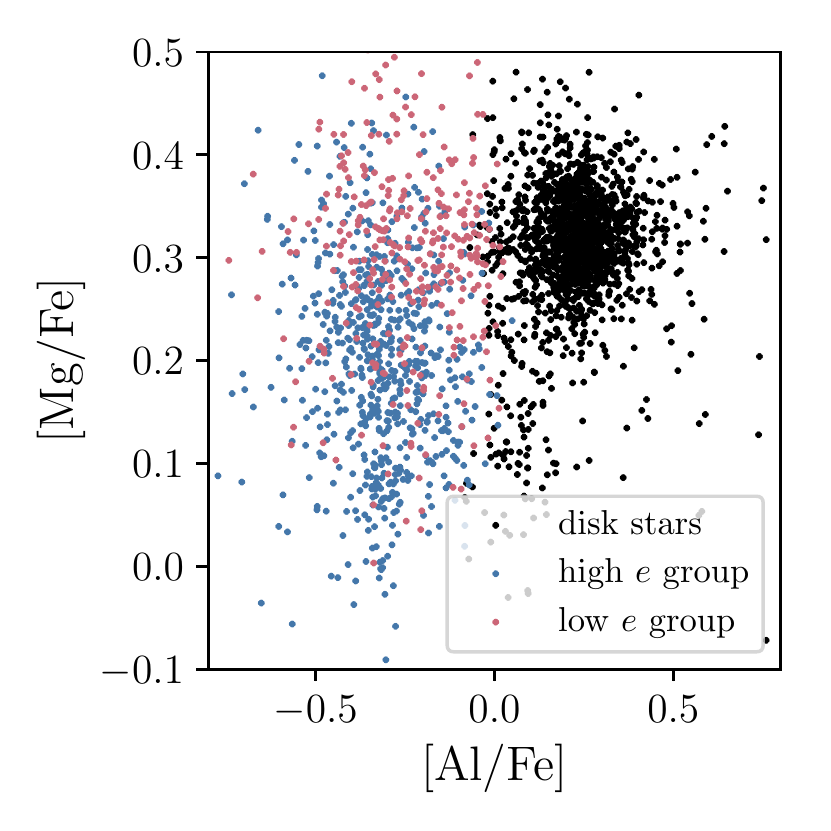}
\caption{\label{fig:mgal}  Distribution of the stars in the different
$k$-means groups in the \mgfe{}-\alfe{} plane.  High $e$ stars
occupy preferentially the low \alfe{}, low \mgfe{} region of the plot, a locus which was identified
in \citet{2015MNRAS.453..758H} as being common to accreted halo populations. Low~$e$ group stars
tend to be more distributed at higher \mgfe{} and slightly higher
\alfe{}. Disc stars (in black) are enhanced in Al and, to a lesser
extent, Mg, relative to the high and low~$e$ groups.}
\end{figure}

We show the eccentricity distributions of the $k$-means groups in
Figure \ref{fig:eccdist}. The high and low~\afe{} disc groups are
combined into a single group in this plot, as these stars
are not of interest to this work, except as a comparison sample.
The halo stars naturally split up into two groups, one
with intermediate eccentricities, peaking at $e \sim 0.5$, and
another with very high eccentricities, peaking at $e \sim
0.9$. The lower $e$ population distribution is fairly broad whereas
the high~$e$ population is very strongly peaked toward the highest
$e$ values (with some small skew to low $e$). The high and low~$e$
groups contain 679 and 318 stars, respectively. We also show the
$k$-means groups in \alfe{}-\mgfe{} space in Figure \ref{fig:mgal}.
This abundance plane was shown by \citet{2015MNRAS.453..758H} to
discriminate well between accreted and in-situ stellar populations
in the Milky Way, with the former occupying the low \alfe{}, low
\mgfe{} region of the plot.  Indeed, we find that the halo population
separates out from the disc population very clearly in this plane,
being characterised by lower Al and Mg abundances.  Moreover, while
there is considerable overlap between high and low~$e$ populations
in this abundance plane, the high~$e$ group occupies a lower \mgfe{}
locus, on average, than the low~$e$ group.

In conclusion, the results above show that once APOGEE chemical
compositions are combined with kinematics inferred from Gaia DR2
proper motions, $\sim 2/3$ of the stars in the accreted halo
population identified in previous studies are characterised by
highly eccentric orbits.  In the following sub-sections we examine
the properties of this population in both orbital and element
abundance space compared to those of their low~$e$ counterparts.

\subsection{Chemical Compositions}

\subsubsection{Metallicity Distributions} \label{sec:mdfs}

\begin{figure}
\includegraphics[width=1.0\columnwidth]{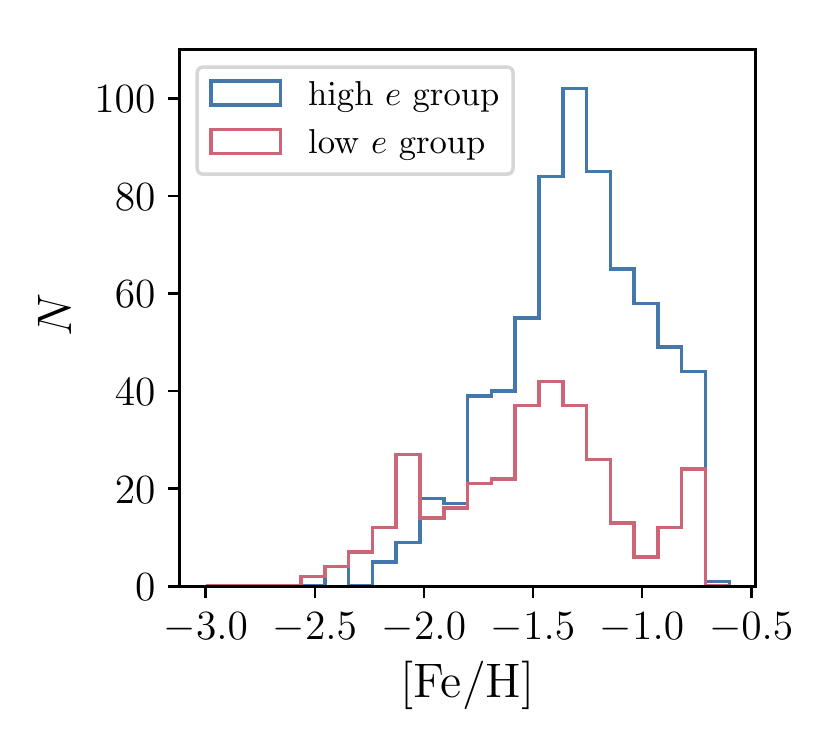}
\caption{\label{fig:mdfs} Raw metallicity distribution functions
of the high and low~$e$ populations.  Unlike their low~$e$ counterparts,
the high $e$ population shows a strong peak at about \feh{}$\sim
-1.3$, resembling MDFs of Local Group dwarfs. The low~$e$ group
shows peaks at both \feh{} $\sim -2.1$ and $\sim -1.4$, and stars
piled up at the imposed upper \feh{} limit at \feh{} $= -0.7$.
}
\end{figure}

In this Section we examine more closely the chemical compositions
of the halo stellar populations, keeping track of the differences
between the high and low~$e$ groups.  We start by examining the
metallicity distribution function (MDF) of these populations.  The
{\it raw} MDFs are displayed in Figure \ref{fig:mdfs}, and are {\it
not} corrected for the APOGEE selection function. Correction for
selection function effects can not be performed to the full sample
of stars studied here, as many of the sample stars belong to magnitude
bins on plates where observations were incomplete in DR14 and so
are not part of the statistical sample \citep[see][for details on
the selection function of APOGEE and the targeting strategy,
respectively]{2014ApJ...790..127B,2013AJ....146...81Z}. 
As a result, a selection function correction would leave us with a
much smaller sample, resulting in more uncertain MDFs, particularly
in the case of the smaller low~$e$ population sample.  We nevertheless
compared the lower $N$, corrected and raw MDFs for the larger
high~$e$ sample, which is statistically more robust, finding them
to be in good overall agreement with similar peak position and
width.  Therefore, we conclude that selection function
effects do not affect substantially our halo sample.  Moreover, it
is reasobable to assume that relative differences between the MDFs
of the high and low~$e$ populations are not affected significantly
by selection biases from the targeting strategy, and that any
differences in the MDFs are likely to be real. Therefore, we can
confidently compare the MDFs of the high and low~$e$ populations,
at least to first order.  It should be noted here that \feh{} is
included in the $k$-means analysis, and so these MDFs are reflective
of the metallicities of the groups as defined by that procedure.

With the above caveats in mind, examining Figure \ref{fig:mdfs} we
learn that the MDF of the high~$e$ population peaks at \feh{}$\sim$--1.3
and has a width of approximately 0.6 dex (FWHM), with a tail towards
the metal-poor end, resembling the MDF of a classical closed-box
model.  Its MDF is in fact qualitatively similar to those of
relatively massive dwarf galaxies in the Local Group
\citep[e.g.][]{2015AJ....149..198R}, and in particular quite similar
to those of the Small Magellanic Cloud \citep[SMC,][]{2014MNRAS.442.1680D}
and Leo I \citep{2013ApJ...779..102K}. This is an interesting result
given that \citet{2018ApJ...852...49H}  showed that their LMg
population (of which the high~$e$ stars are a sub-group) had a star
formation history similar to that of massive dwarfs such
as the LMC prior to its accretion, which if accretedwould have likely been
around present day SMC mass \citep[$\sim 3\times 10^{8}
\mathrm{M_{\odot}}$, e.g.][]{2004ApJ...604..176S,2018MNRAS.478.5017R}
at the time that the LMg population was accreted
\citep{2009IAUS..256...81V,2018arXiv180606038H}.

It is also instructive to compare the MDFs of the high and low~$e$
populations.  The most striking difference is that the MDF of the
low~$e$ population has no strong peak, but rather two smaller (likely
significant) peaks at lower metallicities.  The sharp edge at
\feh{}=--0.7 is due to the \feh{} limit adopted in our $k$-means
grouping, and the smaller peak of the low~$e$ population at
\feh{}$>$--1.0 is probably due to disc contamination.  Interestingly,
the two peaks towards the lower metallicity end happen at similar
metallicities to those previously assigned to the inner and outer
halo populations
\citep[e.g.][]{2007Natur.450.1020C,2014A&A...568A...7A,2015A&A...577A..81F}.
Most importantly, the MDFs of the high and low~$e$ populations are
strikingly different and, as argued above, this difference
is almost certainly insensitive to target selection effects, which
then strongly suggests that the two populations have different
origins.

It is not entirely clear how the difference in shape between the
MDFs of the high and low~$e$ populations can be understood in terms
in of the halo accretion history.  While the MDF of the high~$e$
population lends support to the notion that those stars were injected
into the halo as part of one major accretion event, it is
hard to draw any strong conclusion regarding the low~$e$ population.

\subsubsection{Abundance Ratios}
\label{sec:abundances}
\begin{figure}
\includegraphics[width=1.0\columnwidth]{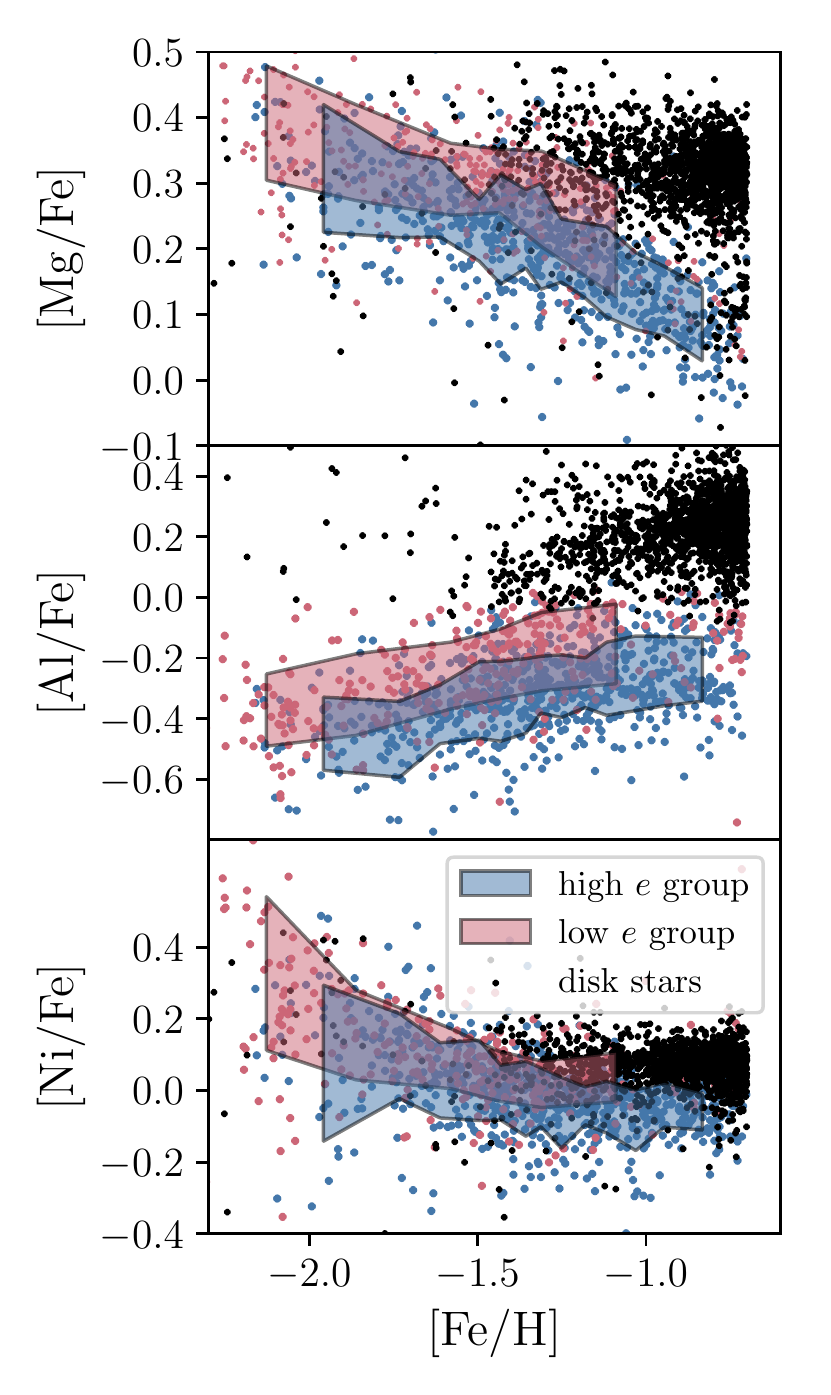}
\caption{\label{fig:mgalni} Abundance ratios for the $k$-means
groups as a function of [Fe/H]. These elements are good representations
of $\alpha$, odd-$Z$ and Fe peak elements that are measured by
APOGEE. The coloured bands show the interquartile range in bins of
50 stars sorted in increasing \feh{}. The high and low~$e$ populations
are characterised by lower Mg and Al than the high-$\alpha$ stars.
The bottom panel shows that the high~$e$ group has slightly lower
[Ni/Fe] than the high-$\alpha$ population, whereas the low~$e$ group
is essentially consistent with the high-$\alpha$ population in
[Ni/Fe]. In all cases, the high~$e$ group median is slightly lower
than the low~$e$ group at fixed \feh{}. The slightly higher \feh{}
of the high~$e$ population is also evident here.  Again, both groups
are depleted in all these elements relative to the disc stars at
that \feh{}.}
\end{figure}

 In this section we inspect abundance ratios of the high and
low~$e$ groups, to gain further insights into the origins of these
two populations.  Figure \ref{fig:mgalni} displays the run of
[Mg/Fe], [Al/Fe], and [Ni/Fe] as a function of \feh{}.  For each
element we take the running median and interquartile range of
$\mathrm{[X/Fe]}$ as a function of \feh{}. We calculate
the median abundance in bins of 50 stars sorted by increasing \feh{},
which is shown by the red and blue coloured bands for the high and
low~$e$ populations, respectively.

The top panel shows that in general, \mgfe{} decreases with \feh{},
and both the high and low~$e$ populations have \mgfe{} lower than
that of the high-$\alpha$ disc stars at same \feh{}, in the metallicity
interval where those populations overlap
($\mathrm{-1.5\simless[Fe/H]\simless-0.5}$), as previously discussed
by \cite{2018ApJ...852...49H} and \citet{2018ApJ...852...50F}.
Moreover, the high and low~$e$ populations occupy slightly separated
sequences, such that the former has lower \mgfe{} at fixed \feh{}.
It appears from the simple running median that the high~$e$ population
exhibits a change in slope of \mgfe{} against \feh{}, at \feh{}$\sim-1.3$,
whilst there is only slight evidence change of slope in the lower
edge of the low~$e$ relation.  In order to test this further, and
to determine more rigorously the location of the change in slope,
we use a Bayesian inference to determine a generative model for the
\mgfe{}-\feh{} distribution of both populations. The full procedure
is outlined in Appendix \ref{sec:appA}, the basic principle being
that we fit a piecewise-linear model to the data of the form:

\begin{equation}
\label{eq:pwlin}
\mathrm{[Mg/Fe]}(\mathrm{[Fe/H]}) =
  \begin{cases}
    m_{1}\mathrm{[Fe/H]} + b_1  &  \quad \mathrm{[Fe/H]} < \mathrm{[Fe/H]}_0\\
    m_{2}\mathrm{[Fe/H]} + b_2  &  \quad \mathrm{[Fe/H]} > \mathrm{[Fe/H]}_0
  \end{cases}
\end{equation}

\noindent where $m_{[1,2]}$ and $b_{[1,2]}$ are the slope and
$y$-intercept either side of the break, which is positioned at
$\mathrm{[Fe/H]}=\mathrm{[Fe/H]}_0$. The fitting procedure accounts
for the error on both \mgfe{} and \feh{}, allowing a proper assessment
of the significance of any fitted breaks. If there is no break in
the range of \mgfe{} and \feh{} considered, the data are fit by a
simple linear relation, as the break is pushed to the edge of the
\feh{} range.

\begin{figure}
\includegraphics[width=1.0\columnwidth]{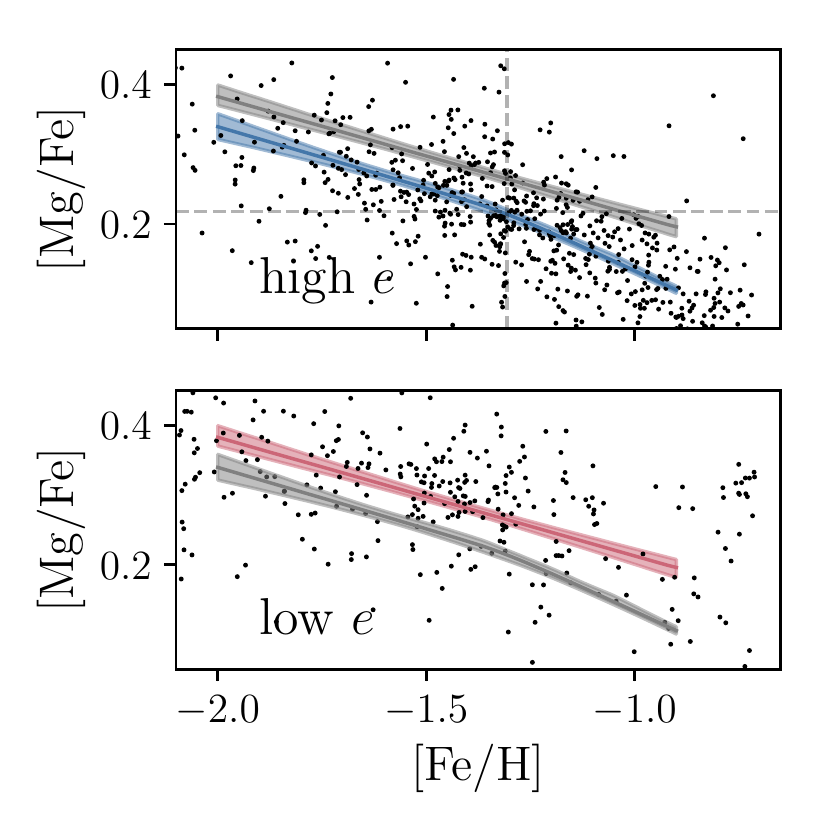}
\caption{\label{fig:mgknee} Best fit models for \mgfe{} as a
function of \feh{} for the high and low~$e$ $k$-means groups (shown
top and bottom, respectively). The best fit model is shown by the
blue and red lines, with the 95$\%$ confidence interval marked by
the banded blue and red regions either side of this line (the fit
for the other group is shown in each panel for reference). The raw
data are shown by the black scatter points. The high~$e$ group is
well-fit by a model with a break at \feh{}$=-1.31^{+0.03}_{-0.06}$
and \mgfe{}$=0.22^{+0.01}_{-0.08}$, as indicated by the dashed
crosshair. The low~$e$ group is better fit by a single linear
relation in the same range of \feh{}. The position of the models
also demonstrates the slightly higher \mgfe{} of the low~$e$ group.}
\end{figure}

We show the models fitted to the high and low~$e$ groups in Figure
\ref{fig:mgknee}. The data for each group are plotted in the black
points, and the best fit and 95$\%$ confidence interval are shown
by the line and coloured bands in each panel. The high~$e$ group
model (shown in the top panel) has a change of slope, indicated by the dashed crosshair, positioned at
\feh{}$=-1.31^{+0.03}_{-0.06}$ and \mgfe{}$=0.22^{+0.01}_{-0.08}$.
The  slope to the metal-poor side of the break is found to
be $-0.15\pm{0.01}$, and is roughly consistent with the slope before
the break in the high \mgfe{} disc star sequence \citep[e.g. that
seen in][]{2015ApJ...808..132H}. A similar metal-poor end slope 
is also present in the work of \citet{2018ApJ...852...50F},
who used APOGEE DR13 data to determine also that a change of slope
is present in low \mgfe{} stars, located at \feh{}$\sim-1.0$ dex,
in rough agreement with that found here. At \feh{}$>\mathrm{[Fe/H]}_0$,
the slope steepens to become $-0.26\pm{0.01}$. The low~$e$ population
is well fit in this range of \feh{} by a simple linear relation
with a slope equal to $-0.14\pm{0.01}$, with no significant break
found. As it is likely that at least some of the stars at the highest
\feh{}, and at high \mgfe{} in this population are high \afe{} disc
contaminants, we also perform the same fit using stars at \feh{}~$<-1.$
dex, and still find no evidence of a break at the same position as
that found in the high~$e$ group. Fitting only low~$e$ stars with
\feh{}~$>-1.1$ dex, we find that the best fit is a single linear
relation with a slope equal to that found when using the whole
low~$e$ group.

\begin{figure*}
\includegraphics[width=\textwidth]{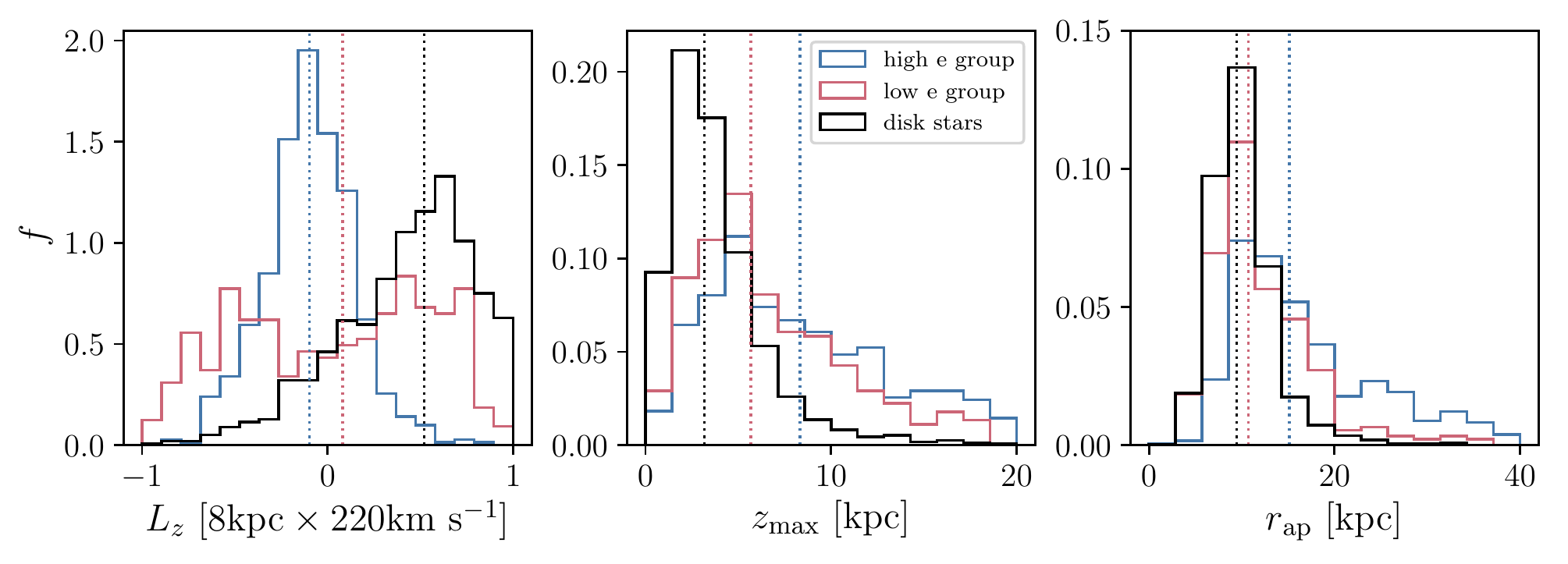} 
\caption{\label{fig:kinematics} Kinematics of the k-means-selected
stars from Figure \ref{fig:characterisation}. The distribution
of azimuthal angular momentum $L_z$ (\emph{left}), maximum vertical
excursion from the disc plane $z_\mathrm{max}$ (\emph{centre}), and
the spherical apocentre radius of orbits $r_\mathrm{ap}$ (\emph{right}),
are shown for the high and low \afe{} disc (red and yellow, respectively)
and the accreted halo population (blue). The halo stars clearly
occupy a very different orbital distribution, having low $L_z$, and
distributions of $z_\mathrm{max}$ and $r_\mathrm{ap}$ that extend
to very large distances. The median $L_z$ is slightly negative,
while the median $z_\mathrm{max}$ and $r_\mathrm{ap}$ are $\sim10$
and $\sim 20$ kpc, respectively. All histograms are normalised such
that the summed probability under \emph{each} group is equal to
unity.}
\end{figure*}

This change of slope in the \mgfe{}-\feh{} relation of the high~$e$
population offers important clues on the nature of this stellar
population. This feature is most commonly interpreted as being due
to the onset of Fe enrichment by SNe Ia.  The \feh{} at which the
change of slope occurs is primarily related to the star formation
efficiency \citep[e.g.][]{2017ApJ...837..183W}, and additionally
depends on the gas inflow and outflow rates of its parent galaxy.
It follows that it should also be a function of the mass of the
parent galaxy, which regulates gas density and gas inflow/outflow.
Data for stellar samples from Local Group dwarf galaxies indicate
the presence of this same change of slope on the Mg-Fe plane, and
it is suggested that the \feh{} at which it occurs is higher in
galaxies with larger luminosity, and presumably higher stellar mass
\citep[see,e.g.][]{2009ARA&A..47..371T}.  Given the \feh{} at which
data for Local Group dwarf galaxies show a change of slope
\citep[see,e.g.][]{2009ARA&A..47..371T}, we estimate that the mass
of the progenitor of this high~$e$ population was somewhere between
$10^{8}$ and $10^{9} M_\odot$. This rough mass estimate is in good
agreement with that inferred from arguments based on the MDF of the
high~$e$ population in Section~\ref{sec:mdfs}.  It moreover places
the mass involved in this accretion event within the range of the
total stellar mass of the Galactic halo \citep[$\sim 4 - 7 \times
10^8\ \mathrm{M_{\odot}}$] {2016ARA&A..54..529B} which argues against
the high~$e$ population belonging to more than one major accretion
event, as multiple accretions at this mass would exceed this mass
limit.

In contrast, the lack of a clear change of slope in the
 distribution of the low~$e$ stars on the Mg-Fe plane, argues
strongly for this group being in fact some mix of stellar populations
with various origins, including stars accreted in different smaller
events, stars formed \emph{in situ}, ejected members of the disc,
 and contamination by members of the low-$e$ tail of the
high~$e$ group.

The middle panel of Figure~\ref{fig:mgalni} shows the run of \alfe{}
with \feh{}.  As pointed out by \citet{2018ApJ...852...49H},
their LMg (and thus also our high~$e$) population is characterised
by much lower \alfe{} than the disc populations at same \feh{}.  In
fact, \alfe{} in those populations is comparable to that
of stars in the Sagittarius dwarf spheroidal \citep{2017ApJ...845..162H}.
 In addition, the high and low~$e$ populations occupy
separate loci in \alfe{}-\feh{} space, with the high $e$
population showing lower \alfe{}.  For both populations \alfe{}
increases with \feh{}. As an odd-$Z$ element, Al is an important
component in the explosive ejecta of massive stars, synthesised in
their carbon and nitrogen hydrostatic burning phases. It is also a
product of Asymptotic Giant Branch (AGB) stellar nucleosynthesis,
particularly at \feh{}$> -0.2$ dex \citep[see,
e.g.][]{2016arXiv160408613A}. Importantly, it is known that odd-$Z$
element yields should be metallicity dependent to some extent due
to their dependence on a high neutron surplus
\citep[e.g.][]{1996snih.book.....A}. This fact may explain the
slight correlation observed between \alfe{} and \feh{} in each group
and also in the disc populations.  The depletion in \alfe{} relative
to the MW disc in both halo groups is has no
straightforward explanation, and we speculate that it may indicate
that AGB stars contributed more importantly to the enrichment of
MW disc ISM.

While the trend in Ni abundances as a function of \feh{} of the
high~$e$ group appears to be roughly similar to that of the low~$e$
group, it is noteworthy that the trend is negative in both groups,
such that \nife{} decreases with increasing \feh{}.  Similarly to
the other elements, Ni is slightly depleted in the high~$e$ group
relative to the low~$e$ group, but this depletion is very small.
At fixed \feh{}, the high~$e$ group has a lower \nife{} than the
disc stars, whereas the low~$e$ group is nearly consistent
with them. 

We also studied the other $\alpha$, odd-$Z$ and Fe peak elements
available in APOGEE. These trends appear to be consistent between
most of the elements. It is noteworthy that in almost all elements,
the low~$e$ group appears to show slightly higher abundance ratios
than the high~$e$ group, at fixed \feh{}, which may be difficult
to explain by invoking lower mass accretion events, unless the star
formation in these stellar populations was very short and intense.
Our discussion of the EAGLE simulations in Section \ref{eagle} sheds
some light on this matter.

\subsection{Kinematics}

Figure \ref{fig:kinematics} shows the distribution of the $k$-means
groups in (from left to right) azimuthal angular-momentum $L_z$
(equivalent to the azimuthal action $J_\phi$), the maximum vertical
excursion of orbits above the midplane $z_\mathrm{max}$, and the
apocentric radius $r_\mathrm{ap}$, which is the maximum orbital
distance from the Galactic centre. The median value of each
distribution is shown by the coloured vertical dashed line. 

It is clear from these plots that the halo groups (again,
shown in red and blue) have a very different orbit distribution
than the disc stars (in black).  The latter are characterised by
highly prograde orbits with relatively low vertical excursions.  In
turn, the two accreted halo populations differ substantially,
particularly in terms of their $L_z$ distributions. The high~$e$
stars all have extremely low angular momentum, such that many of
the stars in this population have negative $L_z$. The lower $e$
population has a large spread in $L_z$, such that many stars are
on prograde orbits while others have negative $L_z$, thus moving
on retrograde orbits. The lack of a distinct single peak in $L_z$
space for the low~$e$ group may indicate again that this group is
in fact a superposition of populations with different origins,
including the debris of smaller mass accreted satellites, whereas
the single, clear peak in $L_z$ for the high~$e$ group supports the
notion that this population is mainly the debris of a single
satellite. The median value of $L_z$ for the high~$e$ stars is $-176
\mathrm{\ km\ s^{-1}\ kpc}$.  A slightly retrograde motion for stars
in the halo at roughly these eccentricities was noted by
\citet{2018arXiv180606038H} and, as they suggest, could be either
an effect of the assumption of the solar motion (which is likely
subject to some systematic uncertainties), or could be a true feature
of this population, which further supports its origin in a single
accreted satellite. We find that a slight retrograde motion of the
high~$e$ stars is present when assuming either the
\citet{2010MNRAS.403.1829S} or the \citet{2005ApJ...629..268H} solar
motion measurements, which differ markedly.

The $z_\mathrm{max}$ distribution shows that the stars belonging
to the high and low~$e$ populations are on orbits that take them
very far above the midplane of the Galaxy, extending to $z_\mathrm{max}>10$
kpc, as expected for halo stars. The median $z_\mathrm{max}$ of the
high~$e$ population is higher than the other two populations, at
8.8 kpc. Similarly, the apocentre radii of these stars extend to
very large distances from the Galactic centre with a median
$r_\mathrm{ap} = 15.1$ kpc.  There appears to be a secondary peak
in the distribution at $r_\mathrm{ap} >\sim25$ kpc, which is roughly
consistent with the ``apocenter pile up'' suggested by
\citet{2018arXiv180510288D}.  Considering that roughly 2/3
of the stars in the halo population are on highly eccentric orbits
($e > 0.8$) which are strongly out of the disc plane ($z_\mathrm{max}
> 10$ kpc), and have a median apocentre radius which is comparable
to that of the accreted halo population analysed by
\citet{2018arXiv180510288D}, it seems reasonable to conclude that
the high~$e$ stars are part of that population.

\begin{figure*} \includegraphics[width=\textwidth]{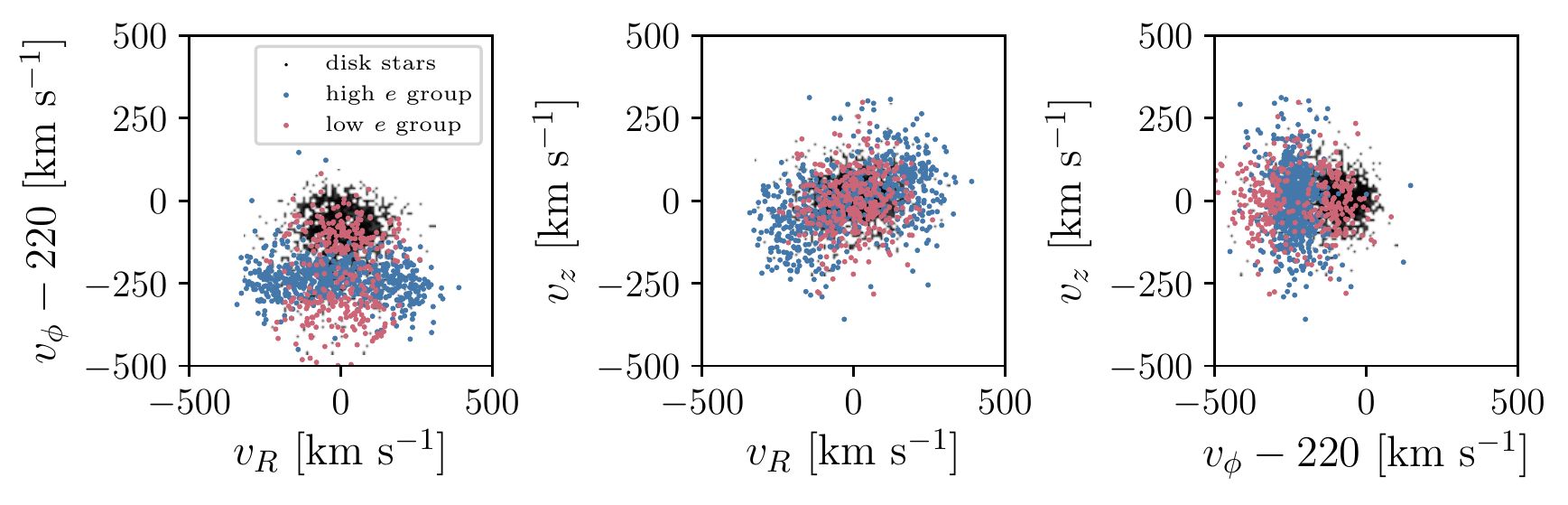}
\caption{\label{fig:polar} Distribution of disc and accreted halo
populations in spherical polar coordinate planes.  On the left
panel, the high $e$ population occupies the same locus in $v_R-v_\phi$
space (left panel) as the population identified by
\citet{2018MNRAS.478..611B}.  The low $e$ population splits into
two populations according to $v_\phi$, with one prograde and one
retrograde component, suggesting that this population may in fact
be from a mix of disc contaminants and debris from smaller satellites.
} \end{figure*}

Further insights into the nature of the high and low~$e$ stellar
populations can be gained by inspection of Figure \ref{fig:polar},
where they are displayed together with disc populations on three
cylindric coordinate velocity planes.  The left panel shows
the $v_R-v_\phi$ plane, where the high $e$ population has a wide
spread in $v_R$ and a much narrower distribution in $v_\phi$.  This
is the same locus as that of the population identified by \citet[][their
Figure 2]{2018MNRAS.478..611B}.  The low $e$ group, on the other
hand, has a clearly bimodal distribution, with two concentrations
above and below the high $e$ population.  The concentration at
higher $v_\phi$ shows {\it prograde} rotation, overlapping well
with the disc population.  The grouping at lower $v_\phi$ is markedly
{\it retrograde}.  These two sub-populations are also clearly seen
on the right panel, where one can notice that their distributions
in $v_z$ are very similar.  The presence of such obvious structure
in velocity space calls for a closer scrutiny of those two
sub-populations in chemical composition space.

We compared the low~$e$ prograde and retrograde populations in MDF,
Mg-Fe, Al-Fe, and Ni-Fe spaces and found them to be, on one hand,
indistinguishable from each other but, on the other hand, clearly
distinct from the high~$e$ population in all diagnostic plots.  It
is likely that there is some degree of inter-contamination between
all these groups, so that the left panel of Figure \ref{fig:polar}
suggests strongly that the prograde population can be contributed
entirely or in part by the disc.  This is also suggested by the
presence of a secondary peak at the high \feh{} end of the low $e$
MDF (see Figure \ref{fig:mdfs}) which is dominated by prograde
stars.  However, quite puzzlingly, the prograde population is
chemically indistinguishable from its retrograde counterpart, which
is unlikely to be contributed by the disc.  We test this further
by re-fitting the relation between \mgfe{} and \feh{} for the
separated prograde and retrograde components, using the procedure
described in Appendix \ref{sec:appA}. We find that in both cases,
the best fit relation matches that found for the combined population
within the uncertainties, and furthermore, find no evidence of a
change in slope in either population.  Without further diagnostics
with which to make a call on the nature of the distinct low $e$
populations, we summarise the situation by concluding that it is
likely to be a mix of stars contributed by contaminants
from the high~$e$ population, smaller accretion events, puffed up
disc, and stars formed {\it in situ}.

\section{Accreted Stellar Populations in EAGLE} \label{eagle}

We have shown that the halo stellar population in the
combined APOGEE and Gaia DR2 samples have MDFs and abundance
patterns that are similar to those of dwarf galaxies in the Local
Group. Furthermore, we found that this population contains
two distinct sub-groups, with different chemical and kinematic
characteristics. One of these groups is characterised by highly
eccentric and slightly retrograde orbits, thus sharing similar
properties with stellar populations ascribed to a major accretion
event in the distant past.  In this Section we examine predictions
by numerical simulations, drawing parallels with the observations
to gain insights into the nature of this accreted halo
population.  The questions we intend to address with this exercise
are the following: is the high~$e$ population the result of a single
relatively massive accretion event, as claimed by other groups?  Do
the simulations provide insight into the nature of the low~$e$
population? Is it composed of a collection of small accretion events,
or is it predominantly formed {\it in situ}? 

The EAGLE simulations are a useful tool to address these questions,
as they provide a  cosmologically motivated history of satellite
accretion for Milky Way-like galaxies, simulated self-consistently
in a cosmological context.  Therefore,  we expect the  $z=0$ dynamical
state of accreted systems around Milky Way analogues in the simulations
to be a good approximation to the observations in the Milky Way
halo. In order to attempt to shed light on the problem using EAGLE, we track star particles formed in galaxies which merged onto disc galaxies that eventually reach a
virial mass roughly equal to that of the Milky Way ($\sim 10^{12}
\mathrm{M_{\odot}}$). At $z=0$, we measure the orbital eccentricity
$e$ of the star particles resulting from all well-resolved accretion
events (i.e., those having $>20$ star particles) of 
systems with stellar mass $M_*$, occuring at redshift $z_\mathrm{merge}$.

\begin{figure}
\includegraphics[width=\columnwidth]{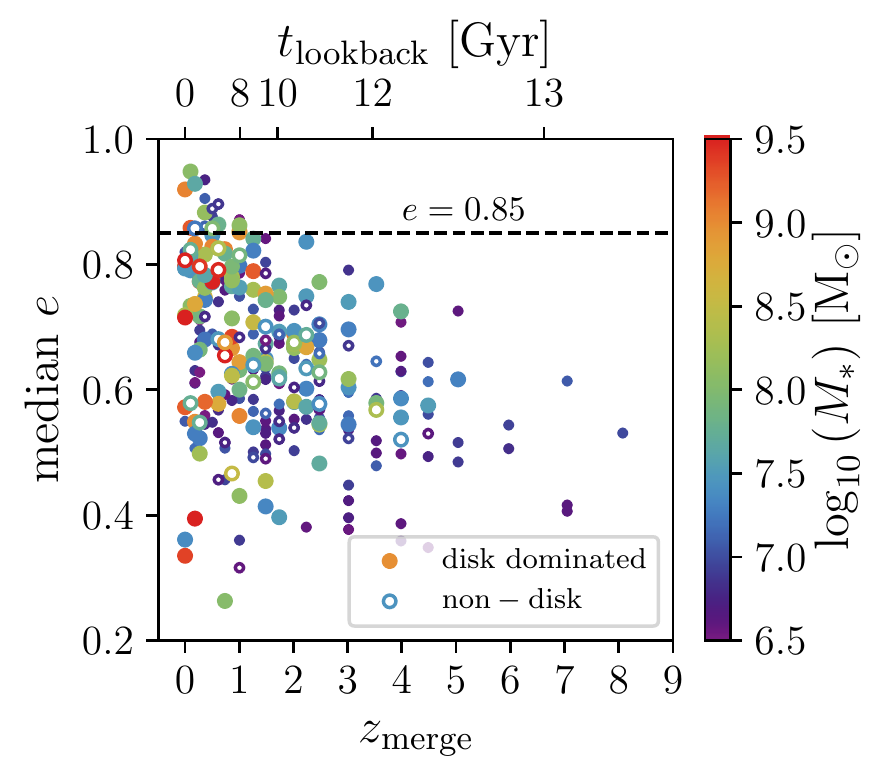}
\caption{\label{fig:eagle}Merger redshift and lookback
time ($z_\mathrm{merge}$ and $t_{\mathrm{lookback}})$ of 296
satellite galaxies accreted onto 22 Milky Way mass haloes in EAGLE
against the median eccentricity $e$ of their stellar debris at
$z=0$. The points are coloured by the stellar mass of the accreted
galaxies at the time of the merger, $M_*$. The larger points denote
those galaxies which have $N > 100$ star particles in EAGLE, whereas
the smaller points are those galaxies with $20 < N < 100$ particles.
Open points indicate satellites accreted onto galaxies which do not
have a clear disk component at $z=0$ (defined as described in Section
\ref{eagledata}). The dashed horizontal line indicates $e = 0.85$,
which is the median eccentricity of the high~$e$ group characterised
in Section \ref{highe}. Only the latest merged haloes have debris
at the highest and lowest $e$, whereas early mergers tend to occupy
intermediate $e$. The most massive mergers also occur later, with
early mergers dominated by low mass haloes.}
\end{figure}

In Figure \ref{fig:eagle},
we show the median $e$ of star particle orbits against
$z_\mathrm{merge}$ for accreted satellites onto the 22
simulated galaxies with Milky Way-like virial mass ($\sim 10^{12}
\mathrm{M_{\odot}}$) in the EAGLE L025N752-Recal simulation.  We
colour each point  according to the stellar mass reached
by the progenitor population prior to merging onto the main
progenitor. The median eccentricity of the high~$e$ group
in the Milky Way halo is indicated as a dashed horizontal line. To
ensure that we properly resolve the orbital properties of the stellar
component of the accreted satellites, we only show galaxies with
20 or more star particles, amounting to a total of 296 accreted
satellites. The satellites with $20 < N < 100$ particles are shown
by small symbols, whereas large symbols are adopted
for satellites with $N > 100$.  Satellites accreted onto
galaxies that are not dominated by a disk component (as discussed
in Section \ref{eagledata}), are plotted as open symbols.

There is a striking trend of median $e$ with $z_{\mathrm{merge}}$
in Figure \ref{fig:eagle}. The maximum median $e$ decreases with
lookback time, creating an upper envelope to the distribution of
the data.  As a result, $z=0$ stellar particles resulting from the
earliest accretions typically have median $e\sim0.5$, whereas the
most recently accreted satellites span a wide range of median
$e$, from $\sim 0.3$ up to $> 0.9$.  As we discuss below, this means
that early accreted satellites were accreted onto intermediate to
low $e$ orbits.  Most importantly, this result implies that 
an accreted satellite whose stars at $z=0$ are on very radial orbits is
unlikely to have been accreted before $z\sim1.5$. We
verify that this trend is not due to numerical noise in Appendix
\ref{sec:appB}, by demonstrating that the same trend is realised
in a larger volume simulation (at lower resolution), with a far
larger sample of accreted satellites (see Figure \ref{fig:l50}). We
discuss the implications of this finding for the formation of the
Milky Way thick disc, as suggested by \citet{2018arXiv180606038H},
in Section \ref{finale}.

As expected from hierarchica galaxy mass growth in a
standard $\Lambda$-CDM universe, it can be seen in Figure \ref{fig:eagle}
that, at the highest redshifts, merged satellite galaxies are
relatively low mass.  Conversely, the most massive galaxies
were accreted recently.  Almost all accretion of stellar systems
with $M_* \simgreater 10^9 M_\odot$ happened at $z \simless 1$.
This is mainly due to the time taken for satellites to build up
their stellar mass. The fact that the satellite debris with the
highest median $e$ must have been accreted late also means that
these satellites are likely to be high mass. We find that the median
mass of accreted satellites with median $e>0.85$ is $4.1\times
10^{8}\ \mathrm{M_{\odot}}$, and a minimum mass of $0.77 \times
10^{8}\ \mathrm{M_{\odot}}$. The median accretion time of high $e$
debris is $z \sim 0.6$. This suggests that the high~$e$ Milky Way
halo population we identify is likely to have had  $M_* \gtrsim
10^{8}\ \mathrm{M_{\odot}}$ and have been accreted after $z\sim
1.5$.  Accretion events such as this are relatively rare in EAGLE,
with only 5 of the 22 ($\sim~23\%$) central galaxies having accreted
any system with $10^8 < M_* < 10^9 \mathrm{M_{\odot}}$ after $z=1.5$
\emph{and} retaining a median $e > 0.8$ at $z=0$.  For two of those
five simulated galaxies, the aggregate of all accreted stellar mass
far exceeds that of the Galactic halo (e.g. by a factor of 2 or
more), leaving only 3 galaxies ($\sim$~14\%) with accretion profiles
resembling that suggested by the observations reported in this
paper.  It is also worth noticing that the high~$e$ populations of
the latter three galaxies, although dominated by one massive accretion
event, had not insignificant contributions by accretions of smaller
mass systems.  The latter suggests that it is possible that the
high~$e$ population of the Galactic halo may have contributions by
low mass systems as well. 

Perhaps most importantly, the above discussion indicates that the
accretion event which deposited the high~$e$ Milky Way debris was
quite unusual for a galaxy of its $z=0$ virial mass.  If one further
considers that the Milky Way is currently accreting a similarly
massive galaxy (the Sagittarius Dwarf), and may have accreted
yet another massive system in the distant past (the \emph{Kraken}),
as proposed by \citet{2018MNRAS.tmp.1537K}, our results suggest
that the overall accretion history of the Milky Way has been quite
atypical when compared to the rest of the galaxies of same halo
mass. This is an interesting result in light of the findings reported
by \citet{2018MNRAS.477.5072M}, which suggest that the
distribution of Milky Way disc stars on the \afe{}-\feh{} plane
indicate that it has undergone an unusual accretion history, for
it's stellar mass.  We discuss the implications of that work further
in Section \ref{finale}.

\begin{figure}
\includegraphics[width=\columnwidth]{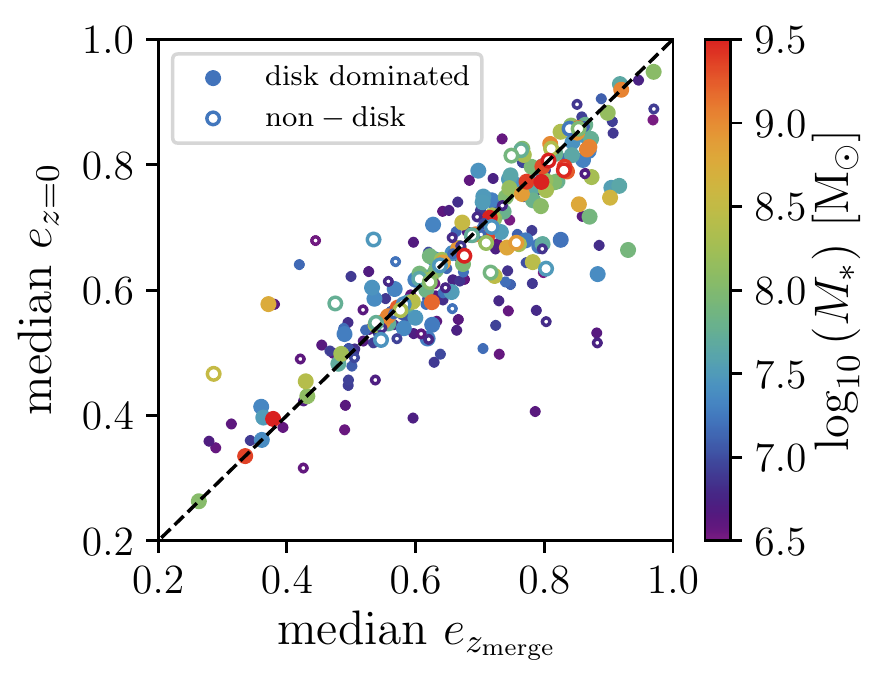}
\caption{\label{fig:echange} Median eccentricities at merger time
$e_{z_\mathrm{merge}}$ against median eccentricity at $z=0$ $e_{z=0}$
for satellites accreted onto Milky Way mass haloes in EAGLE. The
points are coloured by the stellar mass of the accreted galaxies
at the time of the merger, $M_*$.  As in Figure \ref{fig:eagle},
point sizes denote the number of stellar particles in the satellite,
and the open points show satellites accreted onto galaxies with no
clear disk component.  Debris which falls below the dashed unity
line has been circularised following accretion, whereas the debris
above this line has been radialised after accretion. The majority
of debris has a similar $e$ at $z=0$ to the time at which it was
accreted, but a few accreted satellites are circularised/radialised
following accretion. High mass satellite debris has undergone the
least change in median $e$.}
\end{figure}

In Figure \ref{fig:echange}, we show how the eccentricity of satellite
debris changes between the time of the merger and $z=0$. We measure
the initial eccentricity by taking the median eccentricity of the
star particles in the snapshot immediately following that in which
the satellite is identified as being merged onto the main progenitor.
It is clear from this figure that the orbit eccentricities are not
greatly changed following accretion. The median change in eccentricity
is small, at $\sim 4\%$, although a few accreted satellites have
changed their median eccentricity by as much as $\sim 20 \%$. It
is noticeable that the debris whose eccentricity has changed most
is that from lower mass satellites.  As noted previously, these are
the satellites commonly accreted earlier, although we find no
significant trend between $z_\mathrm{merge}$ and the change in
eccentricity, meaning that changes in the median $e$ are not strongly
time dependent.  By the same token, the highest mass satellite
debris are those which have undergone the least change in median
$e$, likely because they were accreted late, when the gravitational
potential of the host dark matter halo is no longer varying
significantly. Furthermore, no satellites accreted at lower $e$
than $\sim 0.7$ have debris that is radialised to higher than $e
\sim 0.8$ by $z=0$. The finding that median $e$ changes very little
is important, as it means that the trend between $z_\mathrm{merge}$
and median $e$ at $z=0$ (Figure~\ref{fig:eagle}) is cosmological
in nature, and not due to dynamical effects, substantiating the
idea that high orbit eccentricity is a good indicator of late-time
accretion.  The trend between $z_\mathrm{merge}$ and median $e$ is
likely a product of the changing merger cross-section with cosmic
time, whereby the gravitational potential of the central halo
dictates the maximum impact parameter for a successful accretion
(as opposed to a ``fly-by'').  Galaxies in the early universe are
smaller and less massive, and so approaching satellites must have
a smaller impact parameter to be accreted -- in fact so small
at very high redshift that very eccentric orbits are essentially
impossible to achieve.  For example, if we assume that at high redshifts ($z\sim$3 to 4), galaxies have small virial radii, $r_\mathrm{vir} \sim
50$~kpc, a high $e$ orbit ($\gtrsim 0.8$) can only be attained by a satellite impacting at $r_{\mathrm{peri}}\simless 5.5$~kpc (assuming that the maximum $r_\mathrm{apo}$ is set by the virial radius).  Conversely, for Milky-Way-like centrals with larger virial radii at $z=0$, closer to $r_\mathrm{vir} \sim
250$~kpc, an impacting galaxy can merge anywhere with $r_\mathrm{peri}\simless 28$ kpc for an
orbit with similar eccentricity, meaning that high eccentricity mergers are more likely at lower redshifts. This logic supports our contention
that the high~$e$ group in the Milky Way must have been a relatively
late accretion event.

It is noteworthy that the morphology of the central galaxy does not
seem to impact the trends seen in Figures~\ref{fig:eagle} or
\ref{fig:echange}, since the galaxies without dominant disc components
(open symbols) trace roughly the same locus as  their disc-dominated
counterparts (filled symbols). This is of interest because it
suggests that the galaxy morphology is generally not necessarily
reflective of the galaxy assembly history in terms of merger time,
mass and orbital properties following the merger, at least on the
relatively long timescales shown in these figures. Moreover,
disc and non-disc galaxies have essentially the same mean and
standard deviations around the identity relation in
Figure~\ref{fig:echange} (to within a fraction of a percent).  This
result suggests that the evolving galaxy morphology has little
effect on the orbital evolution of accreted debris once the satellite
is fully unbound.

Regarding the low~$e$ group, the numerical simulations reinforce
the notion that it may be contributed by a combination of smaller
and/or earlier accretion events. Given that intermediate $e$
populations can be accreted at almost any $z_{\mathrm{merge}}$ in
EAGLE, and that the element abundances of this group show no clear
$\mathrm{[Mg/Fe]}$ change of slope in the \feh{} range observed and
no clear MDF peak at moderately high \feh{}, it is then most likely
that these stars were accreted at earlier times than the high~$e$
stars, and in small satellites. In general, earlier and lower mass
accreted satellites in the simulation show $\alpha$-element
enhancements, due to the fact that these must have formed quickly
(with high SFE) to be accreted at early times.

\begin{figure*}
\includegraphics[width=\textwidth]{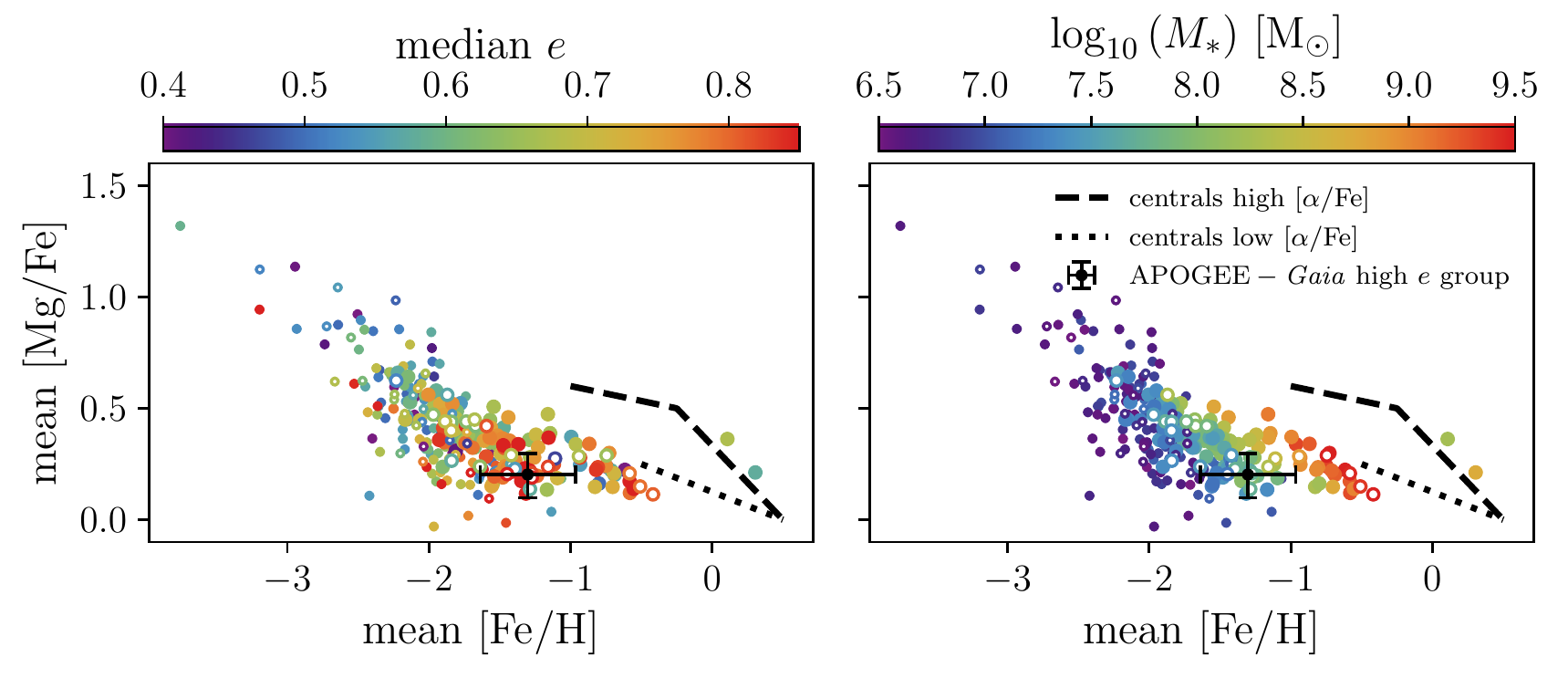}

\caption{\label{fig:eagleabundances} Mean \mgfe{} and \feh{} of the
accreted satellites shown in Figures \ref{fig:eagle} and
\ref{fig:echange}. In both panels, point size and filling
indicate the number of particles of accreted systems, and morphology
of central galaxy, respectively, as in previous figures.  On the
right hand panel, colour denotes the masses of accreted satellites.
The same data are shown in the left hand panel, colour-coded by the
median eccentricity at $z=0$ of the debris.  The dashed and dotted
lines indicate roughly the locus of the disk stars of
MW-like central galaxies in the L025N752-Recal simulation \citep[see,
e.g.][]{2018MNRAS.477.5072M}. It is clear that the abundances of the stars in
the accreted satellites are roughly consistent with those seen in
halo stars in the APOGEE data (the median and 1$\sigma$ scatter of
the high $e$ group is shown by the black error bars).  Less massive
satellites have a large spread in \mgfe, but all have lower \feh{}
than the more massive satellites.  The increasing \feh{} with
satellite mass is very clear. The left hand panel shows that there
is a distinct lack of high eccentricity satellite debris at high
\mgfe{} and low \feh{}, such that, in general, high $e$ debris
occupies a similar locus to that of the high $e$ group in the
observed data.}

\end{figure*}

In Figure \ref{fig:eagleabundances} we plot the {\it mean} \mgfe{}
and \feh{} abundances of the accreted debris shown in Figures
\ref{fig:eagle} and \ref{fig:echange}.  For reference, the dashed
lines display the locus occupied by disc stellar populations of
central galaxies in the MW mass range \citep[For example, galaxies like that
studied in Figure 6 of][]{2018MNRAS.477.5072M}.  In
the right hand panel, we colour the points by the stellar mass of
the accreted satellite, as in the previous figures, as well as
adopting the same size and filling of the points. The left hand
panel shows the same points, but coloured by the median $e$ of the
debris at $z=0$.  The black error bar represents the median
and standard deviation of the stars which are members of
the high $e$ group from the APOGEE-\emph{Gaia} data.

 We first focus on the behaviour of simulated galaxies. It is
immediately clear that the chemical compositions of the EAGLE debris
are broadly consistent with those seen in the APOGEE-\emph{Gaia}
halo stars (Figure~\ref{fig:characterisation}).  The right panel
clearly shows that there is a significant trend between the mass
and mean \feh{} of the accreted satellites, such that more massive
satellites have higher \feh{}.  Comparing the left and right panels,
one concludes that the higher mass satellites, which are characterised
by lower \mgfe{} and higher \feh{}, are those which are more likely
to have higher median $e$.  Very few of the low mass, high \mgfe{}
satellites have median $e \gtrsim 0.7$.  These trends indicate that
we can use both the kinematics and chemistry of the high~$e$
population to constrain the mass of the accreted satellite.  Looking at the
right panel of Figure~\ref{fig:eagleabundances} one would conclude that the
mass would be around $10^8~M_\odot$.  However, the EAGLE simulations
overpredict the metallicity at fixed mass below $10^9~M_\odot$ by about 0.5
dex (Schaye et al. 2015, Fig. 13).  Correcting for that discrepancy the
mass inferred for the satellite associated with the high~$e$ population
would be somewhere between $10^{8.5}~M_\odot$ and $10^9~M_\odot$.

As a further interesting point, we note that the lowest mass satellite
debris in Figure~\ref{fig:eagleabundances} achieve the highest
\mgfe{} abundances.  We interpret this result as being due to the
fact that these systems form quickly with star formation characterised
by short bursts and an early cessation, so that SN Ia could not
contribute substantially to the chemical enrichment of the gas.  As a
caveat, however, we point out that the scatter in \mgfe{} also
increases as the satellites become less massive (and \feh{} decreases).
We emphasise that these low mass satellites are sampled by the
fewest star particles ($N\sim 20$), and so the sampling of the
enrichment from supernovae in EAGLE becomes somewhat stochastic.

We test the numerical convergence of these results in Appendix
\ref{sec:appB}, showing that the result in Figure \ref{fig:eagle}
is robust against changes to the simulation resolution, sub-grid
model and volume. Importantly, we show that degrading the resolution
of the simulation does not significantly change the result shown
in Figure \ref{fig:echange}, which indicates that the simulations
are adequately resolving the interaction of the central galaxy and
the accreted satellite debris. It is important to emphasise that
the time at which the initial $e$ is measured is, of course, an
important factor which affects the conclusions drawn from this
result and its comparison with previous work, which we discuss 
in Section \ref{finale}.

\section{Summary and Conclusions} \label{finale}

This paper presents an analysis of chemical compositions and orbital
information of halo stellar populations, based on APOGEE
and Gaia data.  We applied k-means clustering to identify subgroups
in chemical composition and kinematics space, considering the
abundances of $\alpha$, odd-Z and Fe peak elements, and orbital
eccentricity.  We have shown that $\sim~2/3$ of the accreted halo
stars exhibit very high orbital eccentricity and display chemical
compositions that are characteristic of those seen in massive dwarf
galaxy satellites of the Milky Way today, suggesting that this
population is likely the progeny of a single, massive accretion
event which occurred early in the history of the Milky Way Galaxy, as suggested by other groups
\citep[e.g.][]{2018MNRAS.475.1537M,2018arXiv180510288D,2018arXiv180606038H,2018MNRAS.478..611B}.
The remaining 1/3 of the sample consists of stars with low orbital
eccentricities and slightly higher abundance ratios than the high~$e$
population.  The latter stars likely result from a mixture of
different origins, including the remnants of less massive accretion
events, {\it in situ} star formation, disc heating, and likely some
contamination from the high $e$ population.

We further examine this scenario by studying a numerical simulation
from the EAGLE suite. We demonstrate that satellite galaxies accreted
into MW mass haloes show clear trends between the time of accretion
and the median eccentricity of the accreted stars at $z=0$.
According to the simulation, only satellites accreted at $z\simless
1.5$ result in median debris as high as those of the high~$e$
population identified in our data ($>0.8$). This constraint also
means that such satellites are likely to be accreted at  relatively
high stellar mass ($M_*\gtrsim 10^{8} \mathrm{M_{\odot}}$). 
A stricter constraint is obtained by comparing the median position
of high~$e$ stars with those of EAGLE accreted systems on the Mg-Fe
plane, whereby we infer a mass in the range $10^{8.5}-10^9~M_\odot$.
We also showed that, according to the simulation, the median
eccentricity of accreted debris generally does not appear to evolve
significantly over time, further suggesting that a high orbital
eccentricity is a good indicator of a relatively recent merger.
Analysis of the numerical simulation further suggests that a massive
accretion event such as that identified in the APOGEE/Gaia data is
not very common for a Milky Way-like galaxy, indicating an unusual
accretion history in same vein as suggested in our previous
work \citep{2018MNRAS.477.5072M}.

The high~$e$ population identified in the combined APOGEE/Gaia DR2
sample appears to be the same population as  those discovered by
\citet{2018MNRAS.478..611B}, \citet{2018arXiv180510288D},
\citet{2018arXiv180606038H}, and \citet{2018ApJ...863..113H}.  We
have shown that the kinematics of the high~$e$ group is consistent
with that found in these studies, having very low, slightly negative
mean $L_z$ consistent with the \citet{2018arXiv180606038H} population,
and $r_\mathrm{ap}$ as high as 40 kpc, with a median of 15.1 kpc,
and a suggestive secondary peak at $r_\mathrm{ap} > 20$, in rough
consistency with the \citet{2018arXiv180510288D} population. Our
high~$e$ population has a median eccentricity in good agreement
with those of the populations discussed by \citet{2018MNRAS.478..611B}
and the clusters measured by \citet{2018arXiv180500453M}.

Using cosmological zoom-in simulations, \citet{2018MNRAS.478..611B}
found that the growing discs of central galaxies act to radialise
the orbits or accreted satellite debris as they accrete.
This result is in seeming contradiction with our findings that
debris eccentricity is relatively unchanged after satellite accretion
(Figure~\ref{fig:echange}). However
the $e_{z_\mathrm{merge}}$ we measure in the simulation is that of the
debris once the satellite is fully unbound and merged to the central
galaxy, and therefore likely already having undergone any radialisation
from its initial orbit before accretion.  Our result simply shows that the
orbits do not evolve greatly following accretion onto the galaxy,
and is therefore not in direct contradiction with the work of
\citet{2018MNRAS.478..611B} \citep[or][who showed a similar effect
in an earlier study]{2017MNRAS.464.2882A}. However, the finding
that the galaxies in EAGLE which do not have a significant disc
components at $z=0$ appear to follow the same trends in
$e(z=0)$-$z_\mathrm{merge}$ space (Figure \ref{fig:eagle}) suggests
that disc growth may not be the fundamental factor in driving these
trends.

\citet{2018MNRAS.tmp.1537K} suggest that the age-metallicity
distribution of the Galactic globular cluster population can be
used to infer the formation and assembly history of the Milky Way.
They identify one very massive ($M_* > 10^9\mathrm{M_\odot}$)
accretion event (\emph{Kraken}), which is associated with the
overabundance of metal-rich GCs in the accreted cluster branch of
the age-metallicity relation and has no known debris. They
argue that the \emph{Kraken} is associated with GCs that reside
within 5 kpc of the Galactic center.  \citet{2018MNRAS.tmp.1537K}
further point out that the GCs identified by \citet{2018arXiv180500453M}
(many of which are located at $r_{GC}> 10$ kpc) closely match the sample
they ascribe to the \emph{Canis Major} accretion event.
That would suggest that the stellar population reported in this
paper would by association also be a part of their proposed \emph{Canis
Major} accretion event \citep[as would also be the case of the
system identified by,
e.g.,][]{2018MNRAS.478..611B,2018ApJ...852...49H,2018arXiv180606038H,
2010A&A...511L..10N}. 

In addition to these results based on simulations, it is worth
pointing out that \citet{2013MNRAS.436..122L} associated metal poor
GCs in the Milky Way with an accretion event with mass between $10^8
< M_* < 10^9\ \mathrm{M_\odot}$ (consistent with our estimation for
the high $e$ population progenitor), showing that these clusters
were on very low angular momentum orbits. In summary, all of these
results point towards confirmation that the assembly history of the
Milky Way has been very active, and quite atypical compared to other
galaxies of similar mass.

As mentioned previously, the notion of a massive accretion event
onto the Galaxy is not an entirely new one. The earlier
work of \citet{2010A&A...511L..10N} and \citet{2012A&A...538A..21S}
demonstrated that the halo divides into groups in its $\alpha$-element
abundances, with the kinematics of the lower \afe{} group resembling
that of an accreted population. Even earlier work by
\citet{2003ApJ...585L.125B} showed that a low angular momentum
 stellar population (resembling the one described in this
paper) is present in the data used by \citet{2000AJ....119.2843C}, and possibly also in those upon which \citet{1962ApJ...136..748E}
based their scenario for the formation of the halo.
\citet{2003ApJ...585L.125B} suggested that this population may
indeed be the debris of an accreted dwarf galaxy.  

Finally, \citet{2018arXiv180606038H} suggest that the merger
event reported in this paper and by other groups has occurred
approximately 10~Gyr ago and was responsible for the dynamical
heating responsible for the formation of the thick disc. The results
from analysis of EAGLE suggest that if the debris at $e \sim 0.85$
was accreted in a single satellite, then this is likely to have
occurred around 8-9 Gyr ago (Figure~\ref{fig:eagle}).  The
\citet{2018arXiv180606038H} merger time is based on the minimum
isochronal age of the stars in their sample so it is possibly
indicative of the final time of star formation rather than the
actual accretion time.  On the other hand, EAGLE gives the time at
which the satellite became bound to the central halo, so these
timescales are potentially consistent.  Ages of stars in the Milky
Way high \afe{} disc population are generally found to be older
than or similar to $\sim 10$ Gyr
\citep[e.g.][]{2013A&A...560A.109H,2016MNRAS.456.3655M,2017arXiv170600018M},
so this does suggest that this population was in place before the
merger occurred. It is worth mentioning, however, that analysis of
the origin of $\alpha$-enhanced populations in EAGLE suggest that
these stars must form in an early collapse in a period of rapid gas
accretion, in order to foster the high density ISM necessary to
generate a short enough gas consumption time to consume the high
\afe{} gas into stars before it is polluted by SN Ia
\citep{2018MNRAS.477.5072M}, rather than forming in an
initially thin disc that was later heated. This scenario is
consistent with the thick disc being the result of the geometric
combination of the thick, centrally concentrated high \afe{} disc
and the extended, flared low \afe{} disc
\citep[e.g][]{2017arXiv170600018M,2016arXiv160901168M,2015ApJ...804L...9M}.
Further work on this newly found halo component, and the high \afe{}
disc, will surely shed more light on this discussion.

\section*{Acknowledgements}

The authors thank the anonymous reviewer for a careful and
helpful report. We thank Marie Martig for insightful discussions
during the preparation of this paper, and Robert Crain for crucial
comments on a late draft of the manuscript. We also thank Vasily
Belokurov, Lachlan Lancaster and Ryan Leaman for useful comments
on the originally submitted mansucript. JTM acknowledges an STFC
doctoral studentship. JP gratefully acknowledges funding from a
European Research Council consolidator grant (ERC-CoG-646928-Multi-Pop).
CRH acknowledges the NSF Graduate Research Fellowship through grant
DGE-1315231. JB received support from the Natural Sciences and
Engineering Research Council of Canada (NSERC; funding reference
number RGPIN-2015-05235). JB also received partial support from an
Alfred P. Sloan Fellowship. CAP is thankful to the Spanish Government
for funding for his research through program AYA2017-86389-P. PBT
acknowledges  Fondecyt-Conicyt Regular 1150334. This project was
developed in part at the 2018 Gaia Sprint, hosted by the Center for
Computational Astrophysics of the Flatiron Institute in New York
City. This research made use of the cross-match service provided
by CDS, Strasbourg. Analyses and plots presented in this article
used \texttt{iPython}, and packages in the \texttt{SciPy} ecosystem
\citep{Jones:2001aa,4160265,4160251,5725236}.  The study made use
of high performance computing facilities at Liverpool John Moores
University, partly funded by the Royal Society and LJMU's Faculty
of Engineering and Technology..We acknowledge the Virgo Consortium
for making their simulation data available.  The eagle simulations
were performed using the DiRAC-2 facility at Durham, managed by the
ICC, and the PRACE facility Curie based in France at TGCC, CEA,
Bruy\`eres- le-Chatel.

This work has made use of data from the European Space Agency (ESA)
mission Gaia (~\url{http://www.cosmos.esa.int/gaia}), processed by
the Gaia Data Processing and Analysis Consortium (DPAC,
~\url{http://www.cosmos.esa.int/web/gaia/dpac/consortium}). Funding
for the DPAC has been provided by national institutions, in particular
the institutions participating in the Gaia Multilateral Agreement.

Funding for the Sloan Digital Sky Survey IV has been provided by the Alfred P. Sloan Foundation, the U.S. Department of Energy Office of Science, and the Participating Institutions. SDSS-IV acknowledges
support and resources from the Center for High-Performance Computing at
the University of Utah. The SDSS web site is www.sdss.org.

SDSS-IV is managed by the Astrophysical Research Consortium for the 
Participating Institutions of the SDSS Collaboration including the 
Brazilian Participation Group, the Carnegie Institution for Science, 
Carnegie Mellon University, the Chilean Participation Group, the French Participation Group, Harvard-Smithsonian Center for Astrophysics, 
Instituto de Astrof\'isica de Canarias, The Johns Hopkins University, 
Kavli Institute for the Physics and Mathematics of the Universe (IPMU) / 
University of Tokyo, the Korean Participation Group, Lawrence Berkeley National Laboratory, 
Leibniz Institut f\"ur Astrophysik Potsdam (AIP),  
Max-Planck-Institut f\"ur Astronomie (MPIA Heidelberg), 
Max-Planck-Institut f\"ur Astrophysik (MPA Garching), 
Max-Planck-Institut f\"ur Extraterrestrische Physik (MPE), 
National Astronomical Observatories of China, New Mexico State University, 
New York University, University of Notre Dame, 
Observat\'ario Nacional / MCTI, The Ohio State University, 
Pennsylvania State University, Shanghai Astronomical Observatory, 
United Kingdom Participation Group,
Universidad Nacional Aut\'onoma de M\'exico, University of Arizona, 
University of Colorado Boulder, University of Oxford, University of Portsmouth, 
University of Utah, University of Virginia, University of Washington, University of Wisconsin, 
Vanderbilt University, and Yale University.




\bibliographystyle{mnras}
\bibliography{bib} 

\begin{thebibliography}{}
\makeatletter
\relax
\def\mn@urlcharsother{\let\do\@makeother \do\$\do\&\do\#\do\^\do\_\do\%\do\~}
\def\mn@doi{\begingroup\mn@urlcharsother \@ifnextchar [ {\mn@doi@}
  {\mn@doi@[]}}
\def\mn@doi@[#1]#2{\def\@tempa{#1}\ifx\@tempa\@empty \href
  {http://dx.doi.org/#2} {doi:#2}\else \href {http://dx.doi.org/#2} {#1}\fi
  \endgroup}
\def\mn@eprint#1#2{\mn@eprint@#1:#2::\@nil}
\def\mn@eprint@arXiv#1{\href {http://arxiv.org/abs/#1} {{\tt arXiv:#1}}}
\def\mn@eprint@dblp#1{\href {http://dblp.uni-trier.de/rec/bibtex/#1.xml}
  {dblp:#1}}
\def\mn@eprint@#1:#2:#3:#4\@nil{\def\@tempa {#1}\def\@tempb {#2}\def\@tempc
  {#3}\ifx \@tempc \@empty \let \@tempc \@tempb \let \@tempb \@tempa \fi \ifx
  \@tempb \@empty \def\@tempb {arXiv}\fi \@ifundefined
  {mn@eprint@\@tempb}{\@tempb:\@tempc}{\expandafter \expandafter \csname
  mn@eprint@\@tempb\endcsname \expandafter{\@tempc}}}

\bibitem[\protect\citeauthoryear{{Abolfathi} et~al.,}{{Abolfathi}
  et~al.}{2018}]{2018ApJS..235...42A}
{Abolfathi} B.,  et~al., 2018, \mn@doi [\apjs] {10.3847/1538-4365/aa9e8a},
  \href {http://adsabs.harvard.edu/abs/2018ApJS..235...42A} {235, 42}

\bibitem[\protect\citeauthoryear{{Ahn} et~al.,}{{Ahn}
  et~al.}{2012}]{2012ApJS..203...21A}
{Ahn} C.~P.,  et~al., 2012, \mn@doi [\apjs] {10.1088/0067-0049/203/2/21}, \href
  {http://adsabs.harvard.edu/abs/2012ApJS..203...21A} {203, 21}

\bibitem[\protect\citeauthoryear{{Allende Prieto} et~al.,}{{Allende Prieto}
  et~al.}{2014}]{2014A&A...568A...7A}
{Allende Prieto} C.,  et~al., 2014, \mn@doi [\aap]
  {10.1051/0004-6361/201424053}, \href
  {http://adsabs.harvard.edu/abs/2014A%26A...568A...7A} {568, A7}

\bibitem[\protect\citeauthoryear{{Amorisco}}{{Amorisco}}{2017}]{2017MNRAS.464.2882A}
{Amorisco} N.~C.,  2017, \mn@doi [\mnras] {10.1093/mnras/stw2229}, \href
  {http://adsabs.harvard.edu/abs/2017MNRAS.464.2882A} {464, 2882}

\bibitem[\protect\citeauthoryear{{Andrews}, {Weinberg}, {Sch{\"o}nrich}  \&
  {Johnson}}{{Andrews} et~al.}{2017}]{2016arXiv160408613A}
{Andrews} B.~H.,  {Weinberg} D.~H.,  {Sch{\"o}nrich} R.,   {Johnson} J.~A.,
  2017, \mn@doi [\apj] {10.3847/1538-4357/835/2/224}, \href
  {http://adsabs.harvard.edu/abs/2017ApJ...835..224A} {835, 224}

\bibitem[\protect\citeauthoryear{{Arnett}}{{Arnett}}{1996}]{1996snih.book.....A}
{Arnett} D.,  1996, {Supernovae and Nucleosynthesis: An Investigation of the
  History of Matter from the Big Bang to the Present}.
Princeton Press

\bibitem[\protect\citeauthoryear{{Belokurov}, {Erkal}, {Evans}, {Koposov}  \&
  {Deason}}{{Belokurov} et~al.}{2018}]{2018MNRAS.478..611B}
{Belokurov} V.,  {Erkal} D.,  {Evans} N.~W.,  {Koposov} S.~E.,   {Deason}
  A.~J.,  2018, \mn@doi [\mnras] {10.1093/mnras/sty982}, \href
  {http://adsabs.harvard.edu/abs/2018MNRAS.478..611B} {478, 611}

\bibitem[\protect\citeauthoryear{{Binney}}{{Binney}}{2012}]{2012MNRAS.426.1324B}
{Binney} J.,  2012, \mn@doi [\mnras] {10.1111/j.1365-2966.2012.21757.x}, \href
  {http://adsabs.harvard.edu/abs/2012MNRAS.426.1324B} {426, 1324}

\bibitem[\protect\citeauthoryear{{Binney} \& {Tremaine}}{{Binney} \&
  {Tremaine}}{2008}]{2008gady.book.....B}
{Binney} J.,  {Tremaine} S.,  2008, {Galactic Dynamics: Second Edition}.
Princeton University Press

\bibitem[\protect\citeauthoryear{{Bird}, {Kazantzidis}, {Weinberg}, {Guedes},
  {Callegari}, {Mayer}  \& {Madau}}{{Bird} et~al.}{2013}]{2013ApJ...773...43B}
{Bird} J.~C.,  {Kazantzidis} S.,  {Weinberg} D.~H.,  {Guedes} J.,  {Callegari}
  S.,  {Mayer} L.,   {Madau} P.,  2013, \mn@doi [\apj]
  {10.1088/0004-637X/773/1/43}, \href
  {http://adsabs.harvard.edu/abs/2013ApJ...773...43B} {773, 43}

\bibitem[\protect\citeauthoryear{{Bland-Hawthorn} \&
  {Gerhard}}{{Bland-Hawthorn} \& {Gerhard}}{2016}]{2016ARA&A..54..529B}
{Bland-Hawthorn} J.,  {Gerhard} O.,  2016, \mn@doi [\araa]
  {10.1146/annurev-astro-081915-023441}, \href
  {http://ukads.nottingham.ac.uk/abs/2016ARA%26A..54..529B} {54, 529}

\bibitem[\protect\citeauthoryear{{Blanton} et~al.,}{{Blanton}
  et~al.}{2017}]{2017AJ....154...28B}
{Blanton} M.~R.,  et~al., 2017, \mn@doi [\aj] {10.3847/1538-3881/aa7567}, \href
  {http://adsabs.harvard.edu/abs/2017AJ....154...28B} {154, 28}

\bibitem[\protect\citeauthoryear{{Bovy}}{{Bovy}}{2015}]{2015ApJS..216...29B}
{Bovy} J.,  2015, \mn@doi [\apjs] {10.1088/0067-0049/216/2/29}, \href
  {http://adsabs.harvard.edu/abs/2015ApJS..216...29B} {216, 29}

\bibitem[\protect\citeauthoryear{{Bovy}, {Rix}  \& {Hogg}}{{Bovy}
  et~al.}{2012a}]{2012ApJ...751..131B}
{Bovy} J.,  {Rix} H.-W.,   {Hogg} D.~W.,  2012a, \mn@doi [\apj]
  {10.1088/0004-637X/751/2/131}, \href
  {http://adsabs.harvard.edu/abs/2012ApJ...751..131B} {751, 131}

\bibitem[\protect\citeauthoryear{{Bovy}, {Rix}, {Liu}, {Hogg}, {Beers}  \&
  {Lee}}{{Bovy} et~al.}{2012b}]{2012ApJ...753..148B}
{Bovy} J.,  {Rix} H.-W.,  {Liu} C.,  {Hogg} D.~W.,  {Beers} T.~C.,   {Lee}
  Y.~S.,  2012b, \mn@doi [\apj] {10.1088/0004-637X/753/2/148}, \href
  {http://adsabs.harvard.edu/abs/2012ApJ...753..148B} {753, 148}

\bibitem[\protect\citeauthoryear{{Bovy}, {Rix}, {Hogg}, {Beers}, {Lee}  \&
  {Zhang}}{{Bovy} et~al.}{2012c}]{2012ApJ...755..115B}
{Bovy} J.,  {Rix} H.-W.,  {Hogg} D.~W.,  {Beers} T.~C.,  {Lee} Y.~S.,   {Zhang}
  L.,  2012c, \mn@doi [\apj] {10.1088/0004-637X/755/2/115}, \href
  {http://adsabs.harvard.edu/abs/2012ApJ...755..115B} {755, 115}

\bibitem[\protect\citeauthoryear{{Bovy} et~al.,}{{Bovy}
  et~al.}{2014}]{2014ApJ...790..127B}
{Bovy} J.,  et~al., 2014, \mn@doi [\apj] {10.1088/0004-637X/790/2/127}, \href
  {http://adsabs.harvard.edu/abs/2014ApJ...790..127B} {790, 127}

\bibitem[\protect\citeauthoryear{{Bovy}, {Rix}, {Schlafly}, {Nidever},
  {Holtzman}, {Shetrone}  \& {Beers}}{{Bovy}
  et~al.}{2016}]{2016ApJ...823...30B}
{Bovy} J.,  {Rix} H.-W.,  {Schlafly} E.~F.,  {Nidever} D.~L.,  {Holtzman}
  J.~A.,  {Shetrone} M.,   {Beers} T.~C.,  2016, \mn@doi [\apj]
  {10.3847/0004-637X/823/1/30}, \href
  {http://adsabs.harvard.edu/abs/2016ApJ...823...30B} {823, 30}

\bibitem[\protect\citeauthoryear{{Brook}, {Kawata}, {Gibson}  \&
  {Flynn}}{{Brook} et~al.}{2003}]{2003ApJ...585L.125B}
{Brook} C.~B.,  {Kawata} D.,  {Gibson} B.~K.,   {Flynn} C.,  2003, \mn@doi
  [\apjl] {10.1086/374306}, \href
  {http://adsabs.harvard.edu/abs/2003ApJ...585L.125B} {585, L125}

\bibitem[\protect\citeauthoryear{{Brook}, {Kawata}, {Gibson}  \&
  {Freeman}}{{Brook} et~al.}{2004}]{2004ApJ...612..894B}
{Brook} C.~B.,  {Kawata} D.,  {Gibson} B.~K.,   {Freeman} K.~C.,  2004, \mn@doi
  [\apj] {10.1086/422709}, \href
  {http://adsabs.harvard.edu/abs/2004ApJ...612..894B} {612, 894}

\bibitem[\protect\citeauthoryear{{Carollo} et~al.,}{{Carollo}
  et~al.}{2007}]{2007Natur.450.1020C}
{Carollo} D.,  et~al., 2007, \mn@doi [\nat] {10.1038/nature06460}, \href
  {http://adsabs.harvard.edu/abs/2007Natur.450.1020C} {450, 1020}

\bibitem[\protect\citeauthoryear{{Casagrande} et~al.,}{{Casagrande}
  et~al.}{2016}]{2016MNRAS.455..987C}
{Casagrande} L.,  et~al., 2016, \mn@doi [\mnras] {10.1093/mnras/stv2320}, \href
  {http://adsabs.harvard.edu/abs/2016MNRAS.455..987C} {455, 987}

\bibitem[\protect\citeauthoryear{{Chiba} \& {Beers}}{{Chiba} \&
  {Beers}}{2000}]{2000AJ....119.2843C}
{Chiba} M.,  {Beers} T.~C.,  2000, \mn@doi [\aj] {10.1086/301409}, \href
  {https://ui.adsabs.harvard.edu/#abs/2000AJ....119.2843C} {119, 2843}

\bibitem[\protect\citeauthoryear{{Correa}, {Schaye}, {Clauwens}, {Bower},
  {Crain}, {Schaller}, {Theuns}  \& {Thob}}{{Correa}
  et~al.}{2017}]{2017arXiv170406283C}
{Correa} C.~A.,  {Schaye} J.,  {Clauwens} B.,  {Bower} R.~G.,  {Crain} R.~A.,
  {Schaller} M.,  {Theuns} T.,   {Thob} A.~C.~R.,  2017, \mn@doi [\mnras]
  {10.1093/mnrasl/slx133}, \href
  {http://adsabs.harvard.edu/abs/2017MNRAS.472L..45C} {472, L45}

\bibitem[\protect\citeauthoryear{{Crain} et~al.,}{{Crain}
  et~al.}{2015}]{2015MNRAS.450.1937C}
{Crain} R.~A.,  et~al., 2015, \mn@doi [\mnras] {10.1093/mnras/stv725}, \href
  {http://adsabs.harvard.edu/abs/2015MNRAS.450.1937C} {450, 1937}

\bibitem[\protect\citeauthoryear{{Cullen} \& {Dehnen}}{{Cullen} \&
  {Dehnen}}{2010}]{2010MNRAS.408..669C}
{Cullen} L.,  {Dehnen} W.,  2010, \mn@doi [\mnras]
  {10.1111/j.1365-2966.2010.17158.x}, \href
  {http://adsabs.harvard.edu/abs/2010MNRAS.408..669C} {408, 669}

\bibitem[\protect\citeauthoryear{{Deason}, {Belokurov}, {Evans}  \&
  {Johnston}}{{Deason} et~al.}{2013}]{2013ApJ...763..113D}
{Deason} A.~J.,  {Belokurov} V.,  {Evans} N.~W.,   {Johnston} K.~V.,  2013,
  \mn@doi [\apj] {10.1088/0004-637X/763/2/113}, \href
  {http://adsabs.harvard.edu/abs/2013ApJ...763..113D} {763, 113}

\bibitem[\protect\citeauthoryear{{Deason}, {Belokurov}, {Koposov}  \&
  {Lancaster}}{{Deason} et~al.}{2018}]{2018arXiv180510288D}
{Deason} A.~J.,  {Belokurov} V.,  {Koposov} S.~E.,   {Lancaster} L.,  2018,
  preprint, \href {http://adsabs.harvard.edu/abs/2018arXiv180510288D} {}
  (\mn@eprint {arXiv} {1805.10288})

\bibitem[\protect\citeauthoryear{{Dobbie}, {Cole}, {Subramaniam}  \&
  {Keller}}{{Dobbie} et~al.}{2014}]{2014MNRAS.442.1680D}
{Dobbie} P.~D.,  {Cole} A.~A.,  {Subramaniam} A.,   {Keller} S.,  2014, \mn@doi
  [\mnras] {10.1093/mnras/stu926}, \href
  {http://adsabs.harvard.edu/abs/2014MNRAS.442.1680D} {442, 1680}

\bibitem[\protect\citeauthoryear{{Durier} \& {Dalla Vecchia}}{{Durier} \&
  {Dalla Vecchia}}{2012}]{2012MNRAS.419..465D}
{Durier} F.,  {Dalla Vecchia} C.,  2012, \mn@doi [\mnras]
  {10.1111/j.1365-2966.2011.19712.x}, \href
  {http://adsabs.harvard.edu/abs/2012MNRAS.419..465D} {419, 465}

\bibitem[\protect\citeauthoryear{{Eggen}, {Lynden-Bell}  \& {Sandage}}{{Eggen}
  et~al.}{1962}]{1962ApJ...136..748E}
{Eggen} O.~J.,  {Lynden-Bell} D.,   {Sandage} A.~R.,  1962, \mn@doi [\apj]
  {10.1086/147433}, \href {http://adsabs.harvard.edu/abs/1962ApJ...136..748E}
  {136, 748}

\bibitem[\protect\citeauthoryear{{Eisenstein} et~al.,}{{Eisenstein}
  et~al.}{2011}]{2011AJ....142...72E}
{Eisenstein} D.~J.,  et~al., 2011, \mn@doi [\aj] {10.1088/0004-6256/142/3/72},
  \href {http://adsabs.harvard.edu/abs/2011AJ....142...72E} {142, 72}

\bibitem[\protect\citeauthoryear{{Fern{\'a}ndez-Alvar}
  et~al.,}{{Fern{\'a}ndez-Alvar} et~al.}{2015}]{2015A&A...577A..81F}
{Fern{\'a}ndez-Alvar} E.,  et~al., 2015, \mn@doi [\aap]
  {10.1051/0004-6361/201425455}, \href
  {http://adsabs.harvard.edu/abs/2015A%26A...577A..81F} {577, A81}

\bibitem[\protect\citeauthoryear{{Fern{\'a}ndez-Alvar}
  et~al.,}{{Fern{\'a}ndez-Alvar} et~al.}{2018a}]{2018arXiv180707269F}
{Fern{\'a}ndez-Alvar} E.,  et~al., 2018a, preprint, \href
  {http://adsabs.harvard.edu/abs/2018arXiv180707269F} {} (\mn@eprint {arXiv}
  {1807.07269})

\bibitem[\protect\citeauthoryear{{Fern{\'a}ndez-Alvar}
  et~al.,}{{Fern{\'a}ndez-Alvar} et~al.}{2018b}]{2018ApJ...852...50F}
{Fern{\'a}ndez-Alvar} E.,  et~al., 2018b, \mn@doi [\apj]
  {10.3847/1538-4357/aa9ced}, \href
  {http://adsabs.harvard.edu/abs/2018ApJ...852...50F} {852, 50}

\bibitem[\protect\citeauthoryear{{Ferrero} et~al.,}{{Ferrero}
  et~al.}{2017}]{2017MNRAS.464.4736F}
{Ferrero} I.,  et~al., 2017, \mn@doi [\mnras] {10.1093/mnras/stw2691}, \href
  {http://adsabs.harvard.edu/abs/2017MNRAS.464.4736F} {464, 4736}

\bibitem[\protect\citeauthoryear{{Foreman-Mackey}, {Hogg}, {Lang}  \&
  {Goodman}}{{Foreman-Mackey} et~al.}{2013}]{2013PASP..125..306F}
{Foreman-Mackey} D.,  {Hogg} D.~W.,  {Lang} D.,   {Goodman} J.,  2013, \mn@doi
  [\pasp] {10.1086/670067}, \href
  {http://adsabs.harvard.edu/abs/2013PASP..125..306F} {125, 306}

\bibitem[\protect\citeauthoryear{{Furlong} et~al.,}{{Furlong}
  et~al.}{2017}]{2017MNRAS.465..722F}
{Furlong} M.,  et~al., 2017, \mn@doi [\mnras] {10.1093/mnras/stw2740}, \href
  {http://adsabs.harvard.edu/abs/2017MNRAS.465..722F} {465, 722}

\bibitem[\protect\citeauthoryear{{Gaia Collaboration}, {Brown}, {Vallenari},
  {Prusti}, {de Bruijne}, {Babusiaux}  \& {Bailer-Jones}}{{Gaia Collaboration}
  et~al.}{2018}]{2018arXiv180409365G}
{Gaia Collaboration} {Brown} A.~G.~A.,  {Vallenari} A.,  {Prusti} T.,  {de
  Bruijne} J.~H.~J.,  {Babusiaux} C.,   {Bailer-Jones} C.~A.~L.,  2018,
  preprint, \href {http://adsabs.harvard.edu/abs/2018arXiv180409365G} {}
  (\mn@eprint {arXiv} {1804.09365})

\bibitem[\protect\citeauthoryear{{Garc{\'{\i}}a P{\'e}rez}
  et~al.,}{{Garc{\'{\i}}a P{\'e}rez} et~al.}{2016}]{2016AJ....151..144G}
{Garc{\'{\i}}a P{\'e}rez} A.~E.,  et~al., 2016, \mn@doi [\aj]
  {10.3847/0004-6256/151/6/144}, \href
  {http://adsabs.harvard.edu/abs/2016AJ....151..144G} {151, 144}

\bibitem[\protect\citeauthoryear{Goodman \& Weare}{Goodman \&
  Weare}{2010}]{goodmanweare2010}
Goodman J.,  Weare J.,  2010, Comm. App. Math. and Comp. Sci., 65

\bibitem[\protect\citeauthoryear{{Gunn} et~al.,}{{Gunn}
  et~al.}{2006}]{2006AJ....131.2332G}
{Gunn} J.~E.,  et~al., 2006, \mn@doi [\aj] {10.1086/500975}, \href
  {http://adsabs.harvard.edu/abs/2006AJ....131.2332G} {131, 2332}

\bibitem[\protect\citeauthoryear{{Harris}}{{Harris}}{1996}]{1996AJ....112.1487H}
{Harris} W.~E.,  1996, \mn@doi [\aj] {10.1086/118116}, \href
  {http://adsabs.harvard.edu/abs/1996AJ....112.1487H} {112, 1487}

\bibitem[\protect\citeauthoryear{{Hasselquist} et~al.,}{{Hasselquist}
  et~al.}{2017}]{2017ApJ...845..162H}
{Hasselquist} S.,  et~al., 2017, \mn@doi [\apj] {10.3847/1538-4357/aa7ddc},
  \href {http://adsabs.harvard.edu/abs/2017ApJ...845..162H} {845, 162}

\bibitem[\protect\citeauthoryear{{Hawkins}, {Jofr{\'e}}, {Masseron}  \&
  {Gilmore}}{{Hawkins} et~al.}{2015}]{2015MNRAS.453..758H}
{Hawkins} K.,  {Jofr{\'e}} P.,  {Masseron} T.,   {Gilmore} G.,  2015, \mn@doi
  [\mnras] {10.1093/mnras/stv1586}, \href
  {http://adsabs.harvard.edu/abs/2015MNRAS.453..758H} {453, 758}

\bibitem[\protect\citeauthoryear{{Hayden} et~al.,}{{Hayden}
  et~al.}{2015}]{2015ApJ...808..132H}
{Hayden} M.~R.,  et~al., 2015, \mn@doi [\apj] {10.1088/0004-637X/808/2/132},
  \href {http://adsabs.harvard.edu/abs/2015ApJ...808..132H} {808, 132}

\bibitem[\protect\citeauthoryear{{Hayes} et~al.,}{{Hayes}
  et~al.}{2018}]{2018ApJ...852...49H}
{Hayes} C.~R.,  et~al., 2018, \mn@doi [\apj] {10.3847/1538-4357/aa9cec}, \href
  {http://adsabs.harvard.edu/abs/2018ApJ...852...49H} {852, 49}

\bibitem[\protect\citeauthoryear{{Haywood}, {Di Matteo}, {Lehnert}, {Katz}  \&
  {G{\'o}mez}}{{Haywood} et~al.}{2013}]{2013A&A...560A.109H}
{Haywood} M.,  {Di Matteo} P.,  {Lehnert} M.~D.,  {Katz} D.,   {G{\'o}mez} A.,
  2013, \mn@doi [\aap] {10.1051/0004-6361/201321397}, \href
  {http://adsabs.harvard.edu/abs/2013A%26A...560A.109H} {560, A109}

\bibitem[\protect\citeauthoryear{{Haywood}, {Di Matteo}, {Lehnert}, {Snaith},
  {Khoperskov}  \& {G{\'o}mez}}{{Haywood} et~al.}{2018}]{2018ApJ...863..113H}
{Haywood} M.,  {Di Matteo} P.,  {Lehnert} M.~D.,  {Snaith} O.,  {Khoperskov}
  S.,   {G{\'o}mez} A.,  2018, \mn@doi [\apj] {10.3847/1538-4357/aad235}, \href
  {http://adsabs.harvard.edu/abs/2018ApJ...863..113H} {863, 113}

\bibitem[\protect\citeauthoryear{{Helmi}, {Babusiaux}, {Koppelman}, {Massari},
  {Veljanoski}  \& {Brown}}{{Helmi} et~al.}{2018}]{2018arXiv180606038H}
{Helmi} A.,  {Babusiaux} C.,  {Koppelman} H.~H.,  {Massari} D.,  {Veljanoski}
  J.,   {Brown} A.~G.~A.,  2018, preprint, \href
  {http://adsabs.harvard.edu/abs/2018arXiv180606038H} {} (\mn@eprint {arXiv}
  {1806.06038})

\bibitem[\protect\citeauthoryear{{Hogg}, {Blanton}, {Roweis}  \&
  {Johnston}}{{Hogg} et~al.}{2005}]{2005ApJ...629..268H}
{Hogg} D.~W.,  {Blanton} M.~R.,  {Roweis} S.~T.,   {Johnston} K.~V.,  2005,
  \mn@doi [\apj] {10.1086/431572}, \href
  {http://adsabs.harvard.edu/abs/2005ApJ...629..268H} {629, 268}

\bibitem[\protect\citeauthoryear{{Hogg}, {Bovy}  \& {Lang}}{{Hogg}
  et~al.}{2010}]{2010arXiv1008.4686H}
{Hogg} D.~W.,  {Bovy} J.,   {Lang} D.,  2010, preprint, \href
  {http://adsabs.harvard.edu/abs/2010arXiv1008.4686H} {} (\mn@eprint {arXiv}
  {1008.4686})

\bibitem[\protect\citeauthoryear{{Holtzman} et~al.,}{{Holtzman}
  et~al.}{2015}]{2015AJ....150..148H}
{Holtzman} J.~A.,  et~al., 2015, \mn@doi [\aj] {10.1088/0004-6256/150/5/148},
  \href {http://adsabs.harvard.edu/abs/2015AJ....150..148H} {150, 148}

\bibitem[\protect\citeauthoryear{{Holtzman} et~al.,}{{Holtzman}
  et~al.}{2018}]{holtzdr14}
{Holtzman} J.,  et~al., 2018, AJ

\bibitem[\protect\citeauthoryear{{Hopkins}}{{Hopkins}}{2013}]{2013MNRAS.428.2840H}
{Hopkins} P.~F.,  2013, \mn@doi [\mnras] {10.1093/mnras/sts210}, \href
  {http://adsabs.harvard.edu/abs/2013MNRAS.428.2840H} {428, 2840}

\bibitem[\protect\citeauthoryear{Hunter}{Hunter}{2007}]{4160265}
Hunter J.~D.,  2007, \mn@doi [Computing in Science Engineering]
  {10.1109/MCSE.2007.55}, 9, 90

\bibitem[\protect\citeauthoryear{Jones, Oliphant, Peterson  et~al.}{Jones
  et~al.}{2001}]{Jones:2001aa}
Jones E.,  Oliphant T.,  Peterson P.,   et~al., 2001, {SciPy}: Open source
  scientific tools for {Python}, \url {http://www.scipy.org/}

\bibitem[\protect\citeauthoryear{{J\"onsson} et~al.,}{{J\"onsson}
  et~al.}{2018}]{jonssondr14}
{J\"onsson} H.,  et~al., 2018, AJ

\bibitem[\protect\citeauthoryear{{Kirby}, {Cohen}, {Guhathakurta}, {Cheng},
  {Bullock}  \& {Gallazzi}}{{Kirby} et~al.}{2013}]{2013ApJ...779..102K}
{Kirby} E.~N.,  {Cohen} J.~G.,  {Guhathakurta} P.,  {Cheng} L.,  {Bullock}
  J.~S.,   {Gallazzi} A.,  2013, \mn@doi [\apj] {10.1088/0004-637X/779/2/102},
  \href {http://adsabs.harvard.edu/abs/2013ApJ...779..102K} {779, 102}

\bibitem[\protect\citeauthoryear{{Koppelman}, {Helmi}  \&
  {Veljanoski}}{{Koppelman} et~al.}{2018}]{2018ApJ...860L..11K}
{Koppelman} H.,  {Helmi} A.,   {Veljanoski} J.,  2018, \mn@doi [\apjl]
  {10.3847/2041-8213/aac882}, \href
  {http://adsabs.harvard.edu/abs/2018ApJ...860L..11K} {860, L11}

\bibitem[\protect\citeauthoryear{{Kruijssen}, {Pfeffer}, {Reina-Campos},
  {Crain}  \& {Bastian}}{{Kruijssen} et~al.}{2018}]{2018MNRAS.tmp.1537K}
{Kruijssen} J.~M.~D.,  {Pfeffer} J.~L.,  {Reina-Campos} M.,  {Crain} R.~A.,
  {Bastian} N.,  2018, \mn@doi [\mnras] {10.1093/mnras/sty1609}, \href
  {http://adsabs.harvard.edu/abs/2018MNRAS.tmp.1537K} {}

\bibitem[\protect\citeauthoryear{{Lancaster}, {Koposov}, {Belokurov}, {Evans}
  \& {Deason}}{{Lancaster} et~al.}{2018}]{2018arXiv180704290L}
{Lancaster} L.,  {Koposov} S.~E.,  {Belokurov} V.,  {Evans} N.~W.,   {Deason}
  A.~J.,  2018, preprint, \href
  {http://adsabs.harvard.edu/abs/2018arXiv180704290L} {} (\mn@eprint {arXiv}
  {1807.04290})

\bibitem[\protect\citeauthoryear{{Leaman}, {VandenBerg}  \& {Mendel}}{{Leaman}
  et~al.}{2013}]{2013MNRAS.436..122L}
{Leaman} R.,  {VandenBerg} D.~A.,   {Mendel} J.~T.,  2013, \mn@doi [\mnras]
  {10.1093/mnras/stt1540}, \href
  {http://adsabs.harvard.edu/abs/2013MNRAS.436..122L} {436, 122}

\bibitem[\protect\citeauthoryear{{Lindegren} et~al.,}{{Lindegren}
  et~al.}{2016}]{2016arXiv160904303L}
{Lindegren} L.,  et~al., 2016, preprint, \href
  {http://adsabs.harvard.edu/abs/2016arXiv160904303L} {} (\mn@eprint {arXiv}
  {1609.04303})

\bibitem[\protect\citeauthoryear{{Mackereth} \& {Bovy}}{{Mackereth} \&
  {Bovy}}{2018}]{2018arXiv180202592M}
{Mackereth} J.~T.,  {Bovy} J.,  2018, preprint, \href
  {http://adsabs.harvard.edu/abs/2018arXiv180202592M} {} (\mn@eprint {arXiv}
  {1802.02592})

\bibitem[\protect\citeauthoryear{{Mackereth} et~al.,}{{Mackereth}
  et~al.}{2017}]{2017arXiv170600018M}
{Mackereth} J.~T.,  et~al., 2017, \mn@doi [\mnras] {10.1093/mnras/stx1774},
  \href {http://adsabs.harvard.edu/abs/2017MNRAS.471.3057M} {471, 3057}

\bibitem[\protect\citeauthoryear{{Mackereth}, {Crain}, {Schiavon}, {Schaye},
  {Theuns}  \& {Schaller}}{{Mackereth} et~al.}{2018}]{2018MNRAS.477.5072M}
{Mackereth} J.~T.,  {Crain} R.~A.,  {Schiavon} R.~P.,  {Schaye} J.,  {Theuns}
  T.,   {Schaller} M.,  2018, \mn@doi [\mnras] {10.1093/mnras/sty972}, \href
  {http://adsabs.harvard.edu/abs/2018MNRAS.477.5072M} {477, 5072}

\bibitem[\protect\citeauthoryear{{Majewski} et~al.,}{{Majewski}
  et~al.}{2017}]{2015arXiv150905420M}
{Majewski} S.~R.,  et~al., 2017, \mn@doi [\aj] {10.3847/1538-3881/aa784d},
  \href {http://adsabs.harvard.edu/abs/2017AJ....154...94M} {154, 94}

\bibitem[\protect\citeauthoryear{{Martig}, {Minchev}, {Ness}, {Fouesneau}  \&
  {Rix}}{{Martig} et~al.}{2016a}]{2016arXiv160901168M}
{Martig} M.,  {Minchev} I.,  {Ness} M.,  {Fouesneau} M.,   {Rix} H.-W.,  2016a,
  preprint, \href {http://ukads.nottingham.ac.uk/abs/2016arXiv160901168M} {}
  (\mn@eprint {arXiv} {1609.01168})

\bibitem[\protect\citeauthoryear{{Martig} et~al.,}{{Martig}
  et~al.}{2016b}]{2016MNRAS.456.3655M}
{Martig} M.,  et~al., 2016b, \mn@doi [\mnras] {10.1093/mnras/stv2830}, \href
  {http://adsabs.harvard.edu/abs/2016MNRAS.456.3655M} {456, 3655}

\bibitem[\protect\citeauthoryear{{Masseron} \& {Gilmore}}{{Masseron} \&
  {Gilmore}}{2015}]{2015MNRAS.453.1855M}
{Masseron} T.,  {Gilmore} G.,  2015, \mn@doi [\mnras] {10.1093/mnras/stv1731},
  \href {http://adsabs.harvard.edu/abs/2015MNRAS.453.1855M} {453, 1855}

\bibitem[\protect\citeauthoryear{{McAlpine} et~al.,}{{McAlpine}
  et~al.}{2016}]{2016A&C....15...72M}
{McAlpine} S.,  et~al., 2016, \mn@doi [Astronomy and Computing]
  {10.1016/j.ascom.2016.02.004}, \href
  {http://adsabs.harvard.edu/abs/2016A%26C....15...72M} {15, 72}

\bibitem[\protect\citeauthoryear{{Minchev}, {Martig}, {Streich}, {Scannapieco},
  {de Jong}  \& {Steinmetz}}{{Minchev} et~al.}{2015}]{2015ApJ...804L...9M}
{Minchev} I.,  {Martig} M.,  {Streich} D.,  {Scannapieco} C.,  {de Jong} R.~S.,
    {Steinmetz} M.,  2015, \mn@doi [\apjl] {10.1088/2041-8205/804/1/L9}, \href
  {http://adsabs.harvard.edu/abs/2015ApJ...804L...9M} {804, L9}

\bibitem[\protect\citeauthoryear{{Myeong}, {Evans}, {Belokurov}, {Sanders}  \&
  {Koposov}}{{Myeong} et~al.}{2018a}]{2018arXiv180500453M}
{Myeong} G.~C.,  {Evans} N.~W.,  {Belokurov} V.,  {Sanders} J.~L.,   {Koposov}
  S.~E.,  2018a, preprint, \href
  {http://adsabs.harvard.edu/abs/2018arXiv180500453M} {} (\mn@eprint {arXiv}
  {1805.00453})

\bibitem[\protect\citeauthoryear{{Myeong}, {Evans}, {Belokurov}, {Amorisco}  \&
  {Koposov}}{{Myeong} et~al.}{2018b}]{2018MNRAS.475.1537M}
{Myeong} G.~C.,  {Evans} N.~W.,  {Belokurov} V.,  {Amorisco} N.~C.,   {Koposov}
  S.~E.,  2018b, \mn@doi [\mnras] {10.1093/mnras/stx3262}, \href
  {http://adsabs.harvard.edu/abs/2018MNRAS.475.1537M} {475, 1537}

\bibitem[\protect\citeauthoryear{{Navarro} et~al.,}{{Navarro}
  et~al.}{2017}]{2017arXiv170901040N}
{Navarro} J.~F.,  et~al., 2017, preprint, \href
  {http://adsabs.harvard.edu/abs/2017arXiv170901040N} {} (\mn@eprint {arXiv}
  {1709.01040})

\bibitem[\protect\citeauthoryear{Nelder \& Mead}{Nelder \&
  Mead}{1965}]{doi:10.1093/comjnl/7.4.308}
Nelder J.~A.,  Mead R.,  1965, \mn@doi [The Computer Journal]
  {10.1093/comjnl/7.4.308}, 7, 308

\bibitem[\protect\citeauthoryear{{Nidever} et~al.,}{{Nidever}
  et~al.}{2014}]{2014ApJ...796...38N}
{Nidever} D.~L.,  et~al., 2014, \mn@doi [\apj] {10.1088/0004-637X/796/1/38},
  \href {http://adsabs.harvard.edu/abs/2014ApJ...796...38N} {796, 38}

\bibitem[\protect\citeauthoryear{{Nidever} et~al.,}{{Nidever}
  et~al.}{2015}]{2015AJ....150..173N}
{Nidever} D.~L.,  et~al., 2015, \mn@doi [\aj] {10.1088/0004-6256/150/6/173},
  \href {http://adsabs.harvard.edu/abs/2015AJ....150..173N} {150, 173}

\bibitem[\protect\citeauthoryear{{Nissen} \& {Schuster}}{{Nissen} \&
  {Schuster}}{2010}]{2010A&A...511L..10N}
{Nissen} P.~E.,  {Schuster} W.~J.,  2010, \mn@doi [\aap]
  {10.1051/0004-6361/200913877}, \href
  {http://adsabs.harvard.edu/abs/2010A%26A...511L..10N} {511, L10}

\bibitem[\protect\citeauthoryear{Pedregosa et~al.,}{Pedregosa
  et~al.}{2011}]{scikit-learn}
Pedregosa F.,  et~al., 2011, Journal of Machine Learning Research, 12, 2825

\bibitem[\protect\citeauthoryear{Perez \& Granger}{Perez \&
  Granger}{2007}]{4160251}
Perez F.,  Granger B.~E.,  2007, \mn@doi [Computing in Science Engineering]
  {10.1109/MCSE.2007.53}, 9, 21

\bibitem[\protect\citeauthoryear{{Price}}{{Price}}{2010}]{2010MNRAS.401.1475P}
{Price} D.~J.,  2010, \mn@doi [\mnras] {10.1111/j.1365-2966.2009.15763.x},
  \href {http://adsabs.harvard.edu/abs/2010MNRAS.401.1475P} {401, 1475}

\bibitem[\protect\citeauthoryear{{Queiroz} et~al.,}{{Queiroz}
  et~al.}{2018}]{2018MNRAS.476.2556Q}
{Queiroz} A.~B.~A.,  et~al., 2018, \mn@doi [\mnras] {10.1093/mnras/sty330},
  \href {http://adsabs.harvard.edu/abs/2018MNRAS.476.2556Q} {476, 2556}

\bibitem[\protect\citeauthoryear{{Ross}, {Holtzman}, {Saha}  \&
  {Anthony-Twarog}}{{Ross} et~al.}{2015}]{2015AJ....149..198R}
{Ross} T.~L.,  {Holtzman} J.,  {Saha} A.,   {Anthony-Twarog} B.~J.,  2015,
  \mn@doi [\aj] {10.1088/0004-6256/149/6/198}, \href
  {http://adsabs.harvard.edu/abs/2015AJ....149..198R} {149, 198}

\bibitem[\protect\citeauthoryear{{Rubele} et~al.,}{{Rubele}
  et~al.}{2018}]{2018MNRAS.478.5017R}
{Rubele} S.,  et~al., 2018, \mn@doi [\mnras] {10.1093/mnras/sty1279}, \href
  {http://adsabs.harvard.edu/abs/2018MNRAS.478.5017R} {478, 5017}

\bibitem[\protect\citeauthoryear{{Santiago} et~al.,}{{Santiago}
  et~al.}{2016}]{2016A&A...585A..42S}
{Santiago} B.~X.,  et~al., 2016, \mn@doi [\aap] {10.1051/0004-6361/201323177},
  \href {http://adsabs.harvard.edu/abs/2016A%26A...585A..42S} {585, A42}

\bibitem[\protect\citeauthoryear{{Schaye} et~al.,}{{Schaye}
  et~al.}{2015}]{2015MNRAS.446..521S}
{Schaye} J.,  et~al., 2015, \mn@doi [\mnras] {10.1093/mnras/stu2058}, \href
  {http://adsabs.harvard.edu/abs/2015MNRAS.446..521S} {446, 521}

\bibitem[\protect\citeauthoryear{{Sch{\"o}nrich}, {Binney}  \&
  {Dehnen}}{{Sch{\"o}nrich} et~al.}{2010}]{2010MNRAS.403.1829S}
{Sch{\"o}nrich} R.,  {Binney} J.,   {Dehnen} W.,  2010, \mn@doi [\mnras]
  {10.1111/j.1365-2966.2010.16253.x}, \href
  {http://adsabs.harvard.edu/abs/2010MNRAS.403.1829S} {403, 1829}

\bibitem[\protect\citeauthoryear{{Schuster}, {Moreno}, {Nissen}  \&
  {Pichardo}}{{Schuster} et~al.}{2012}]{2012A&A...538A..21S}
{Schuster} W.~J.,  {Moreno} E.,  {Nissen} P.~E.,   {Pichardo} B.,  2012,
  \mn@doi [\aap] {10.1051/0004-6361/201118035}, \href
  {http://adsabs.harvard.edu/abs/2012A%26A...538A..21S} {538, A21}

\bibitem[\protect\citeauthoryear{{Searle} \& {Zinn}}{{Searle} \&
  {Zinn}}{1978}]{1978ApJ...225..357S}
{Searle} L.,  {Zinn} R.,  1978, \mn@doi [\apj] {10.1086/156499}, \href
  {http://adsabs.harvard.edu/abs/1978ApJ...225..357S} {225, 357}

\bibitem[\protect\citeauthoryear{{Shetrone} et~al.,}{{Shetrone}
  et~al.}{2015}]{2015ApJS..221...24S}
{Shetrone} M.,  et~al., 2015, \mn@doi [\apjs] {10.1088/0067-0049/221/2/24},
  \href {http://adsabs.harvard.edu/abs/2015ApJS..221...24S} {221, 24}

\bibitem[\protect\citeauthoryear{{Springel}}{{Springel}}{2005}]{2005MNRAS.364.1105S}
{Springel} V.,  2005, \mn@doi [\mnras] {10.1111/j.1365-2966.2005.09655.x},
  \href {http://adsabs.harvard.edu/abs/2005MNRAS.364.1105S} {364, 1105}

\bibitem[\protect\citeauthoryear{{Stanimirovi{\'c}}, {Staveley-Smith}  \&
  {Jones}}{{Stanimirovi{\'c}} et~al.}{2004}]{2004ApJ...604..176S}
{Stanimirovi{\'c}} S.,  {Staveley-Smith} L.,   {Jones} P.~A.,  2004, \mn@doi
  [\apj] {10.1086/381869}, \href
  {http://adsabs.harvard.edu/abs/2004ApJ...604..176S} {604, 176}

\bibitem[\protect\citeauthoryear{{Tolstoy}, {Hill}  \& {Tosi}}{{Tolstoy}
  et~al.}{2009}]{2009ARA&A..47..371T}
{Tolstoy} E.,  {Hill} V.,   {Tosi} M.,  2009, \mn@doi [\araa]
  {10.1146/annurev-astro-082708-101650}, \href
  {http://adsabs.harvard.edu/abs/2009ARA%26A..47..371T} {47, 371}

\bibitem[\protect\citeauthoryear{{Trayford} et~al.,}{{Trayford}
  et~al.}{2015}]{2015MNRAS.452.2879T}
{Trayford} J.~W.,  et~al., 2015, \mn@doi [\mnras] {10.1093/mnras/stv1461},
  \href {http://adsabs.harvard.edu/abs/2015MNRAS.452.2879T} {452, 2879}

\bibitem[\protect\citeauthoryear{{Trayford}, {Theuns}, {Bower}, {Crain},
  {Lagos}, {Schaller}  \& {Schaye}}{{Trayford}
  et~al.}{2016}]{2016MNRAS.460.3925T}
{Trayford} J.~W.,  {Theuns} T.,  {Bower} R.~G.,  {Crain} R.~A.,  {Lagos}
  C.~d.~P.,  {Schaller} M.,   {Schaye} J.,  2016, \mn@doi [\mnras]
  {10.1093/mnras/stw1230}, \href
  {http://adsabs.harvard.edu/abs/2016MNRAS.460.3925T} {460, 3925}

\bibitem[\protect\citeauthoryear{{Trayford} et~al.,}{{Trayford}
  et~al.}{2017}]{2017MNRAS.470..771T}
{Trayford} J.~W.,  et~al., 2017, \mn@doi [\mnras] {10.1093/mnras/stx1051},
  \href {http://adsabs.harvard.edu/abs/2017MNRAS.470..771T} {470, 771}

\bibitem[\protect\citeauthoryear{{Weinberg}, {Andrews}  \&
  {Freudenburg}}{{Weinberg} et~al.}{2017}]{2017ApJ...837..183W}
{Weinberg} D.~H.,  {Andrews} B.~H.,   {Freudenburg} J.,  2017, \mn@doi [\apj]
  {10.3847/1538-4357/837/2/183}, \href
  {http://adsabs.harvard.edu/abs/2017ApJ...837..183W} {837, 183}

\bibitem[\protect\citeauthoryear{{White} \& {Frenk}}{{White} \&
  {Frenk}}{1991}]{1991ApJ...379...52W}
{White} S.~D.~M.,  {Frenk} C.~S.,  1991, \mn@doi [\apj] {10.1086/170483}, \href
  {http://adsabs.harvard.edu/abs/1991ApJ...379...52W} {379, 52}

\bibitem[\protect\citeauthoryear{{Wilson} et~al.,}{{Wilson}
  et~al.}{2010}]{2010SPIE.7735E..1CW}
{Wilson} J.~C.,  et~al., 2010, in Ground-based and Airborne Instrumentation for
  Astronomy III. p. 77351C, \mn@doi{10.1117/12.856708}

\bibitem[\protect\citeauthoryear{{York} et~al.,}{{York}
  et~al.}{2000}]{2000AJ....120.1579Y}
{York} D.~G.,  et~al., 2000, \mn@doi [\aj] {10.1086/301513}, \href
  {http://adsabs.harvard.edu/abs/2000AJ....120.1579Y} {120, 1579}

\bibitem[\protect\citeauthoryear{{Zamora} et~al.,}{{Zamora}
  et~al.}{2015}]{2015AJ....149..181Z}
{Zamora} O.,  et~al., 2015, \mn@doi [\aj] {10.1088/0004-6256/149/6/181}, \href
  {http://adsabs.harvard.edu/abs/2015AJ....149..181Z} {149, 181}

\bibitem[\protect\citeauthoryear{{Zasowski} et~al.,}{{Zasowski}
  et~al.}{2013}]{2013AJ....146...81Z}
{Zasowski} G.,  et~al., 2013, \mn@doi [\aj] {10.1088/0004-6256/146/4/81}, \href
  {http://adsabs.harvard.edu/abs/2013AJ....146...81Z} {146, 81}

\bibitem[\protect\citeauthoryear{{Zasowski} et~al.,}{{Zasowski}
  et~al.}{2017}]{2017AJ....154..198Z}
{Zasowski} G.,  et~al., 2017, \mn@doi [\aj] {10.3847/1538-3881/aa8df9}, \href
  {http://adsabs.harvard.edu/abs/2017AJ....154..198Z} {154, 198}

\bibitem[\protect\citeauthoryear{{van der Marel}, {Kallivayalil}  \&
  {Besla}}{{van der Marel} et~al.}{2009}]{2009IAUS..256...81V}
{van der Marel} R.~P.,  {Kallivayalil} N.,   {Besla} G.,  2009, in {Van Loon}
  J.~T.,  {Oliveira} J.~M.,  eds,  IAU Symposium Vol. 256, The Magellanic
  System: Stars, Gas, and Galaxies. pp 81--92 (\mn@eprint {arXiv} {0809.4268}),
  \mn@doi{10.1017/S1743921308028299}

\bibitem[\protect\citeauthoryear{van~der Walt, Colbert  \& Varoquaux}{van~der
  Walt et~al.}{2011}]{5725236}
van~der Walt S.,  Colbert S.~C.,   Varoquaux G.,  2011, \mn@doi [Computing in
  Science Engineering] {10.1109/MCSE.2011.37}, 13, 22

\makeatother
\end{thebibliography}




\appendix
\section{Modelling \mgfe{} as a function of \feh{}}
\label{sec:appA}
In order to test whether a change in slope is found in the relationship
between \mgfe{} and \feh{} in the identified accreted halo groups,
we use a Bayesian inference to fit a piecewise-linear model to the
data. The form of the piecewise-linear function is given in
Equation~\ref{eq:pwlin}. We follow the general procedure outlined
in Section 7 of \citet{2010arXiv1008.4686H} for fitting models to
data with two-dimensional uncertainties. For completeness, we
re-iterate here the mathematics. The best fitting model is found
by maximising the likelihood function for the parameters $O =
[\mathrm{[Fe/H]}_0, \mathrm{[Mg/Fe]}_0, \theta_1, \theta_2]$ given
the data, which we assume here to be of the form
\begin{equation}
\ln{\mathcal{L}(O|\mathrm{[Fe/H]}, \mathrm{[Mg/Fe]})} = K -
\sum^{N}_{i=1}\left (\frac{\Delta_i^2}{2\Sigma^2_i} + \ln |\Sigma^{2}_i|\right)
\end{equation} 
where $\Delta_i^2$ defines the distance between the data-point $i$
and the model, and $\Sigma^2_i$ is the variance orthogonal to the
model, determined by the covariance matric of the data points. $K$
is a normalisation constant, which is not necessary to consider in
the optimisation. We assume uninformative flat priors on $\theta_{[1,2]}$,
and allow $\mathrm{[Fe/H]}_0$ and $\mathrm{[Mg/Fe]}_0$ to be free.
In this case, $\Delta_i^2$ is defined by
\begin{equation}
\label{eq:delta}
\Delta_i^2 = \hat{\bf{v}}^T \bf{Z}_i - \it{b} \cos\theta_{[1,2]}
\end{equation}
where $\bf{Z}_i$ is the column vector made by (\mgfe{},\feh{}$)_i$,
and $\hat{\bf{v}}$ is the unit vector orthogonal to the model:
\begin{equation}
\hat{\bf{v}} =
\frac{1}{\sqrt{1+m_{[1,2]}^2}}\begin{bmatrix}-m_{[1,2]}\\1\end{bmatrix} =
\begin{bmatrix}-\sin\theta_{[1,2]}\\\cos\theta_{[1,2]} \end{bmatrix}
\end{equation}
where $\theta_{[1,2]}$, here and in Equation \eqref{eq:delta}, is
the angle between the linear model at that \feh{} and the x-axis,
which is equal to $\arctan{m_{[1,2]}}$. $\Sigma^2_i$ is then simply
defined as the projection of the data-point's covariance matrix
$\bf{S}_i$ orthogonal to the model at that \feh{}
\begin{equation}
\Sigma^2_i = \hat{\bf{v}}^T \bf{S}_i \hat{\bf{v}}.
\end{equation}
In this case, we assume that the uncertainties on \mgfe{} and \feh{}
are uncorrelated, such that
\begin{equation}
\bf{S}_i = \begin{bmatrix} \delta \mathrm{[Fe/H]}_i & 0 \\ 0 & \delta \mathrm{[Mg/Fe]}_i \end{bmatrix}
\end{equation}
where we use the catalogue values for the uncertainties on \feh{}
and \mgfe{}. We minimise the negative log-likelihood using a downhill
simplex algorithm \citep{doi:10.1093/comjnl/7.4.308}, and use this
optimal solution to initiate an Markov Chain Monte Carlo (MCMC)
sampling of the posterior PDF of the parameters O using an
affine-invariant ensemble MCMC sampler \citep{goodmanweare2010} as
implemented in the python package \texttt{emcee}
\citep{2013PASP..125..306F}. We report the median and standard
deviation of this posterior PDF as our best-fit parameters.

\section{Numerical Convergence Tests}
\label{sec:appB}
We examine in this appendix the effect on the results presented in Figures
\ref{fig:eagle} and \ref{fig:echange} of varying the simulation
resolution, box-size, and subgrid model parameters, in order to
demonstrate that these results are well converged numerically. We
perform the equivalent analysis to that presented for L025N752-Recal
(in the main body of the paper) on lower resolution volumes of the
EAGLE simulations.

\begin{figure}
\includegraphics[width=\columnwidth]{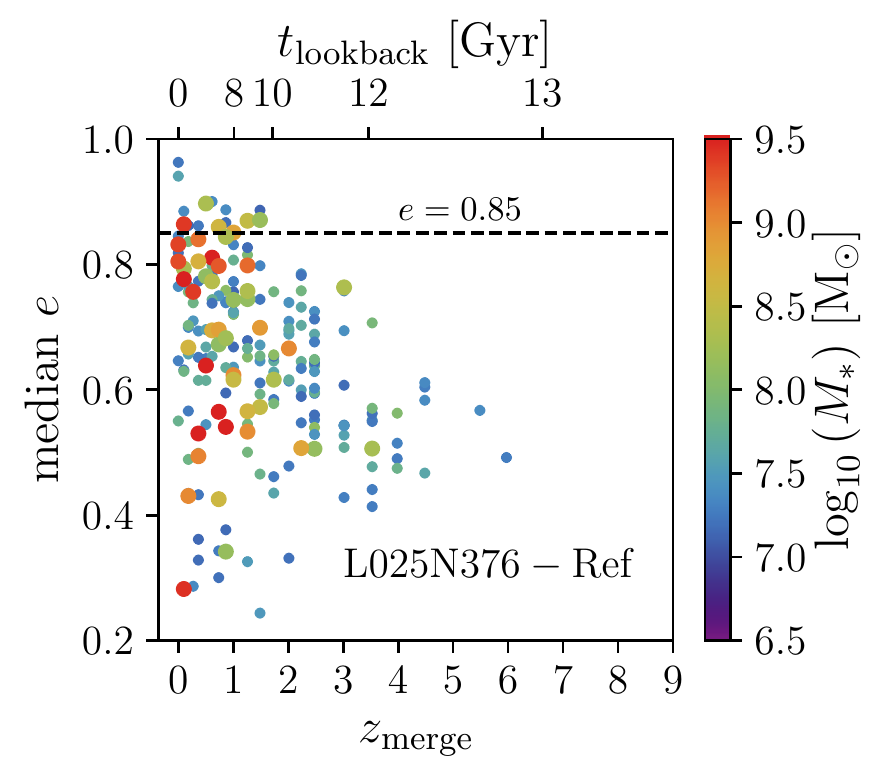}
\caption{\label{fig:n356ref} The merger time $z_\mathrm{merge}$ of
satellites accreted onto Milky Way mass haloes in the L025N356-Ref
simulation against the median eccentricity $e$ of their stellar
debris at $z=0$, produced via an equivalent analysis to that which
produced Figure \ref{fig:eagle}. The trends in Figures \ref{fig:n356ref}
and \ref{fig:eagle} are clearly conserved under an increase in
resolution with no recalibration of the sub-grid feedback model,
demonstrating the strong numerical convergence of these results.}
\end{figure}

\citet{2015MNRAS.446..521S} define `weakly' converged predictions
as those which are unaffected by variation in the simulation
resolution after re-calibrating the sub-grid physics. We test this
by examining the equivalently sized, $L=25$ cMpc, lower resolution
run (at a factor of 10 lower in mass than L025N752-Recal) which
adopts the `Reference' sub-grid model, which we refer to as
L025N376-Ref. We show the resulting equivalent of \ref{fig:eagle}
for L025N356-Ref in \ref{fig:n356ref}. It is clear from this figure
that the general prediction of \ref{fig:eagle} holds, in spite of
the fact that the lower resolution simulation clearly does not
resolve galaxies at masses as low as L025N752-Recal. The upper
envelope of the distribution, which defines the maximum $z=0$
eccentricity of a satellite merged at any given $z$, is still clear
out to $z\sim5$. Given that the sub-grid feedback model is adjusted
between the `Reference' and `Recalibrated' models to maintain the
predictions of the former, from this we gather that the result is
`weakly' converged.

\begin{figure}
\includegraphics[width=\columnwidth]{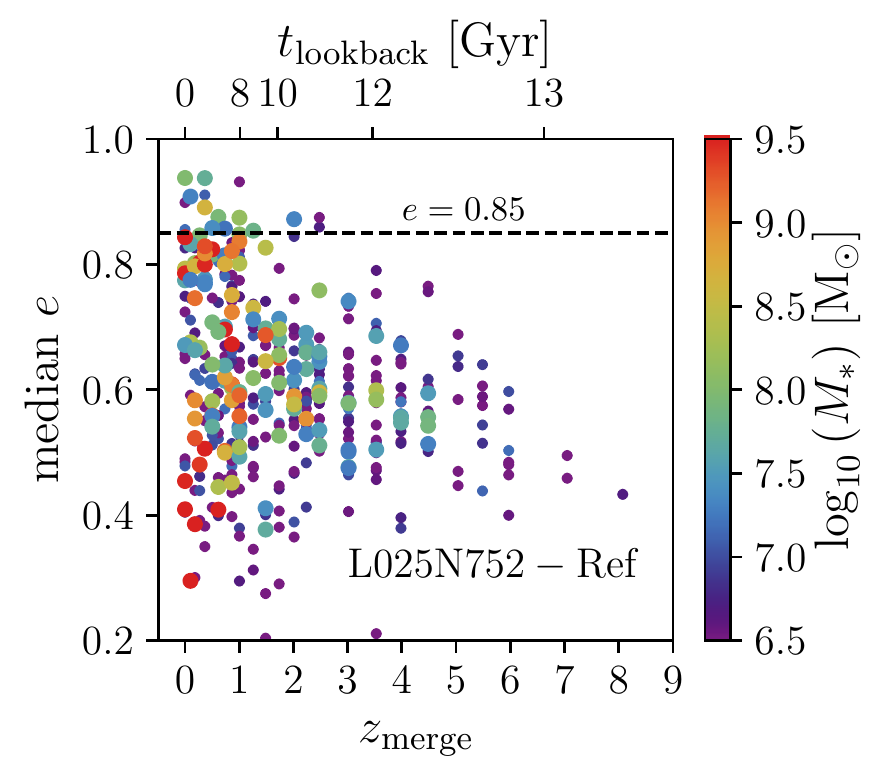}
\caption{\label{fig:n752ref} The merger time $z_\mathrm{merge}$ of
satellites accreted onto Milky Way mass haloes in the L025N752-Ref
simulation against the median eccentricity $e$ of their stellar
debris at $z=0$, produced via an equivalent analysis to that which
produced Figure \ref{fig:eagle}. The upper envelope of $z=0$
eccentricity for a given merger time is reproduced in this lower
resolution simulation, albeit with a different calibration of the
subgrid model, demonstrating the `weak' convergence of this result.}
\end{figure}

In order to check that the prediction of EAGLE for the $z=0$
eccentricity as a function of merger time is `strongly' converged,
we must also ascertain that the trends in Figure \ref{fig:n356ref}
are conserved when the resolution is increased, but the sub-grid
model held fixed. We test this by performing an equivalent analysis
on the L025N752-Ref model, which has a resolution equal to the
L025N752-Recal model, but assumes the same sub-grid physics as
L025N376-Ref. The resulting $z=0$ $e$ against merger time is shown
in Figure \ref{fig:n752ref}. The trends which are seen in both the
low-resolution `Reference' model (Figure \ref{fig:n356ref}) and
high-resolution `Recalibrated' model (Figure \ref{fig:eagle} in the
main text) are clearly conserved here also\footnote{ There is a very slight change in the exact position of the maximum of the distribution as a function of $z_\mathrm{merge}$, however, we contend that this is due to the stochastic sampling of the true underlying distribution.}, demonstrating the
`strong' numerical convergence of these results.

\begin{figure}
\includegraphics[width=\columnwidth]{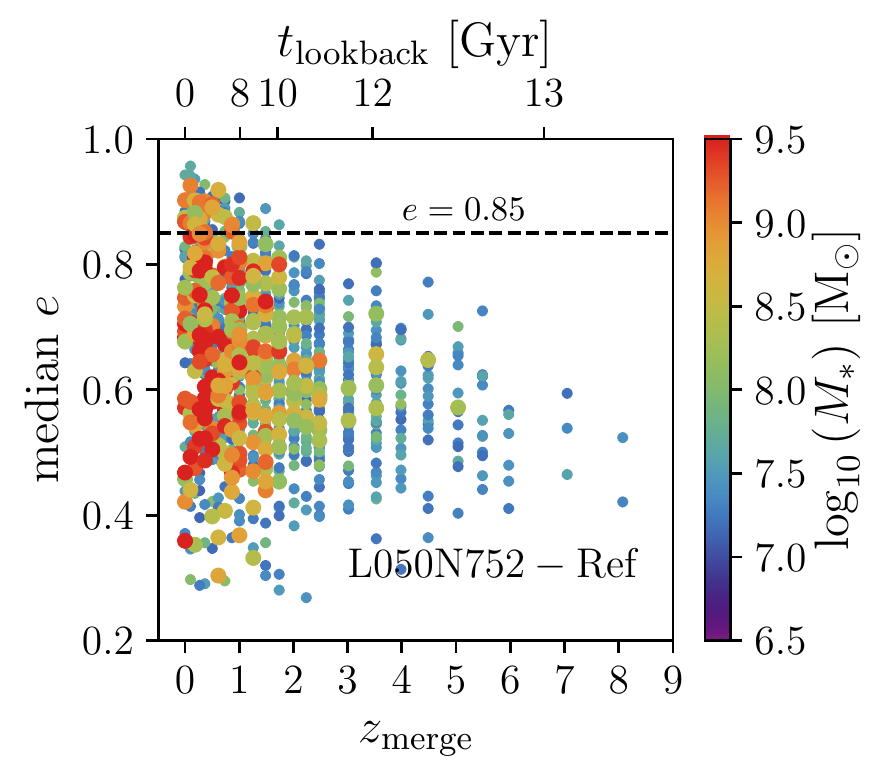}
\caption{\label{fig:l50} The merger time $z_\mathrm{merge}$ against
median eccentricity $e$ of the stellar debris at $z=0$, of the 1154
accreted satellites (with > 20 particles) of 126 central galaxies
from the L050N752-Ref simulation, produced via an equivalent analysis
to that which produced Figure \ref{fig:eagle}. The trends seen in
Figures \ref{fig:eagle}, \ref{fig:n356ref} and \ref{fig:n752ref}
are seen again, even though a much larger sample of accretion events
is analysed.} \end{figure}

We also perform the analysis on a larger volume, $L=50$ cMpc, EAGLE
simulation, which adopts the `Reference' sub-grid model at the same resolution as the smaller volume: L050N752-Ref.
In this much larger volume, we
track the accretion of 1154 satellites onto 126 central haloes,
again with halo masses which are roughly equivalent to that of the
Milky Way. The resulting distribution of satellite debris in median
$e(z=0)$  against merger time is shown in Figure \ref{fig:l50}.
Again, the trend between the maximum $z=0$ eccentricity and
$z_\mathrm{merge}$ is seen, now more clearly, as a result of the
better statistics offered by the increased sample size. This
demonstrates that the results presented from EAGLE are strongly
robust to variations in the simulation resolution, box-size and
sub-grid physics.

\begin{figure}
\includegraphics[width=\columnwidth]{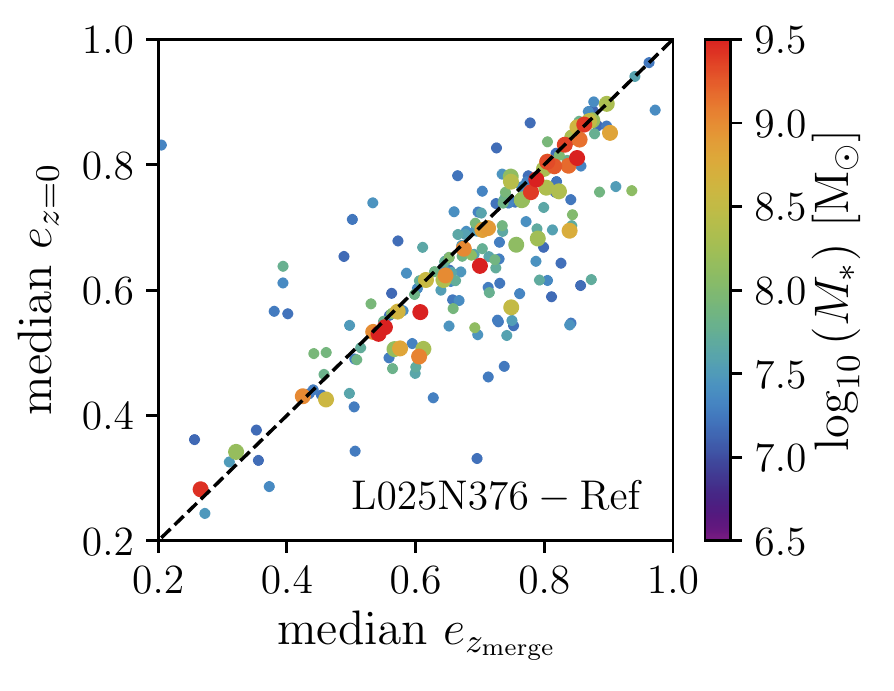}
\caption{\label{fig:echange_lowres} The change in eccentricity between the snapshot immediately after the satellites become unbound and $z=0$ in the L025N376-Ref simulation (equivalent to Figure \ref{fig:echange} in the main text). Degrading the simulation resolution has little effect on this result, which suggests that the lack of any significant radialisation or circularisation of orbits after satellite infall is not due to poorly resolved dynamics, which would act to reduce any scatter upon degradation of the simulation resolution.  }
\end{figure}

Finally, we show in Figure \ref{fig:echange_lowres} that the findings
of Figure \ref{fig:echange} are robust to degradation in the
simulation resolution. It is possible that the dynamical interaction
between the central galaxy and the accreting satellite may be poorly
modelled by the simulations, which are of a relatively low resolution
\citep[as opposed to idealised simulations such as those of
e.g.][]{2017MNRAS.464.2882A}, reducing the action of the central
galaxy on the accreting satellites. By this reasoning, a degradation
in the simulation resolution should decrease any changes to the
orbital properties of the satellites, as these effects would be
worsened. In Figure \ref{fig:echange}, showing the higher resolution
simulation, we find the scatter around the dashed unity line to be
0.13. In the lower resolution we find the scatter to be comparable,
if not slightly increased, at 0.15. As a result, we contend that
these interactions are likely well modelled by the simulation, even
at this relatively low resolution.

\bsp	
\label{lastpage}
\end{document}